\documentclass[twocolumn,a4paper,10pt]{article}

\usepackage{cite}
\usepackage{amsmath,amssymb,amsfonts}
\usepackage{mathtools}
\usepackage{algorithmic}
\usepackage{graphicx}
\DeclareGraphicsExtensions{.pdf,.jpeg,.png}
\usepackage{textcomp}
\usepackage[separate-uncertainty=true,multi-part-units=single,per-mode=symbol]{siunitx}
\usepackage{float}
\usepackage[caption=false,font=footnotesize]{subfig}
\usepackage[utf8]{inputenc} 
\usepackage{authblk}
\usepackage{sectsty}
\sectionfont{\fontsize{11}{12}\selectfont}
\subsectionfont{\fontsize{9}{12}\selectfont}
\usepackage{caption}
\usepackage{times}
\usepackage[a4paper, margin=2cm]{geometry}

\DeclareMathOperator*{\argmin}{arg\,min}

\title{Radio-Coverage-Aware Path Planning\\for Cooperative Autonomous Vehicles}

\date{}

\author[1]{Giuseppe~Baruffa\thanks{Corr. author email: \texttt{giuseppe.baruffa@unipg.it}.\\This work was supported in part by the European Union--Next Generation EU under the Italian National Recovery and Resilience Plan (NRRP), Mission 4, Component 2, Investment 1.3, CUP E83C22004640001, partnership on ``Telecommunications of the Future'' (PE00000001-program ``RESTART'').}}
\author[1]{Luca~Rugini}
\author[1,2]{Francesco~Binucci}
\author[1]{Fabrizio~Frescura}
\author[1]{Paolo~Banelli}
\author[1]{Renzo~Perfetti}
\affil[1]{Department of Engineering, University of Perugia, Perugia, PG 06125, Italy}
\affil[2]{Consorzio Nazionale Interuniversitario per le Telecomunicazioni, Viale G. P. Usberti, 181/A, 43124, Parma, Italy}
\providecommand{\keywords}[1]{\textbf{\small\textit{Index terms---}} {\small #1}}

\let\OLDthebibliography\thebibliography
\renewcommand\thebibliography[1]{
    \OLDthebibliography{#1}
    \setlength{\parskip}{0pt}
    \setlength{\itemsep}{0pt plus 0.3ex}
}

\begin{document}

\maketitle

\begin{abstract}
\small
Fleets of autonomous vehicles (AV) often are at the core of intelligent transportation scenarios for smart cities, and may require a wireless Internet connection to offload computer vision tasks to data centers located either in the edge or the cloud section of the network. Cooperation among AVs is successful when the environment is unknown, or changes dynamically, so as to improve coverage and trip time, and minimize the traveled distance. 
The AVs, while mapping the environment with range-based sensors, move across the wireless coverage areas, with consequences on the experienced access bit rate, latency, and handover rate.
In this paper, we propose to modify the cost of common path planning algorithms such as Dijkstra and A*, so that the best path solution takes into account not only the traveled distance, but also the radio coverage experience. To this aim, several radio-related cost-weighting functions are introduced and tested, to assess the performance of the proposed approaches with extensive simulations. The proposed mapping algorithm can achieve a mapping error probability below $2\%$, while the proposed path-planning algorithms extend the radio coverage of the AVs, with only a limited increase in traveled distance with respect to shortest-path existing methods, such as conventional Dijkstra and A* algorithms.

\end{abstract}

\keywords{Autonomous vehicles, cooperative mapping, path planning, edge computing, task offloading, radio coverage}

\section{Introduction} \label{sec:introduction}

The exploration of an unknown area or geographical region from multiple autonomous vehicles (AV) is well studied in the scientific literature \cite{kumar2023}. 
Multi-AV systems often include robotic and software agents equipped with computing and sensing devices, tailored for smart cities, health care, home services, and intelligent transportation applications \cite{rizk2019}.
The planning of an optimal exploratory path may include multiple objectives, such as coverage, travel time, and path-length. There are several challenges that may complicate the task, such as the presence of complex environments, obstacles, energy or time efficiency constraints, and inter-AV cooperation \cite{kumar2023}. Additionally, the exploration can be focused on separated regions: in this case, coverage must be maximized in each region, and trajectories that connect such regions must be optimized \cite{chen2022}. Multi-AV systems carry additional problems, such as fleet coordination, which needs tailored inter-AV communication protocols to decide optimal trajectories \cite{chen2005}. Cooperative path planning for a group of AVs is indeed required to optimize tasks such as rendezvous, trajectory allocation and region coverage, whether the environment is known or not, and within centralized or decentralized frameworks \cite{zhang2020review}. Before path planning, cooperative mapping should be carried out efficiently, especially if there are communication constraints \cite{paull2015} or real-time requirements \cite{xie2022}.
Additionally, heavy computation burden for computer vision tasks, such as object classification and tracking, could be offloaded to the edge of the network \cite{baruffa2024}, for energy consumption reduction \cite{tan2022} and latency minimization \cite{baruffa2025} purposes.

In this paper we tackle both cooperative mapping and path planning problems. Specifically, we propose suboptimal algorithms that first perform cooperative mapping of an unknown region, and then identify the paths among points of the mapped region. In both cases, we consider the radio coverage limitations possibly experienced by the wireless communication towards the edge computers.
Our main contribution is the definition of radio-related cost functions to be integrated within the cost function of path-planning algorithms such as Dijkstra or A*: more specifically, these new cost functions can be interpreted as radio coverage constraints, that are integrated into trajectory planning by means of Pareto-optimality \cite{hansen2013}. When AVs move between specific points in the known map, the aim is to simultaneously minimize the traveled distance and maximize the radio experience. With respect to the original Dijkstra and A* algorithms, the proposed path planning algorithms are able to significantly increase the experienced radio coverage during travel, with an acceptable penalty in terms of longer traveled distance. The known map is created during a previous cooperative exploration phase with onboard range-based sensors. Nevertheless, path planning and mapping can progress along at the same time. 

The paper organization is as follows. In Section~\ref{sec:litrev} we present a brief review of recent works considering the radio-constrained cooperative mapping problem and the radio-aware path planning problem. As a reference we assume an edge-computing environment, where task offloading is carried out, as outlined in Section~\ref{sec:taskoffload}. The cooperative mapping problem and the associated algorithm we propose, are presented in Section~\ref{sec:coopmap}, whereas the path-planning under radio coverage constraints is investigated in Section~\ref{sec:pathplan}, which also proposes a suboptimal algorithm.
Sections~\ref{sec:rapaplanprob} and \ref{sec:propaplansol} are dedicated to the presentation of the radio-aware path planning problem and of its proposed solution.
Section~\ref{sec:simresults} is devoted to the presentation and discussion of simulation results, while conclusions are drawn in Section~\ref{sec:conclusion}.

In the paper, we adopt the following mathematical notation: $\Vert \mathbf{a} \Vert_l = (\sum_i \vert a_i \vert^l)^{1/l} $ is the entry-wise $l$-norm of a vector ($l=1$ gives the Manhattan norm, $l=2$ gives the Euclidean norm); $\mathbf{B}$ is a matrix with elements indexed by $\mathbf{n}=(n_1,n_2)$, and an element is denoted as $[\mathbf{B}]_{\mathbf{n}} = [\mathbf{B}]_{n_1,n_2}$.

\section{Literature Review} \label{sec:litrev}

This paper focuses mainly on the path-planning problem, that has to be performed using radio coverage information supplied by the network. However, before planning a path, the map of obstacles in a given area must be known or acquired. Thus, we assume that first the AVs cooperate to discover a common map of obstacles. Then, this estimated map is exploited for path planning. Therefore this section briefly reviews some recent works on cooperative mapping and path planning. 

\subsection{Cooperative Mapping}

In \cite{mansouri2018} the authors establish a theoretical framework to map an infrastructure by multiple collaborating agents, by proposing an \textit{a priori} coverage path for each unmanned aerial vehicle (UAV), based on horizontal slices of a 3D space, followed by an off-line cooperative reconstruction of the inspected area.
Since pose graphs are essential for accurate mapping in simultaneous localization and mapping (SLAM), \cite{kim2010} presents a method to jointly optimize the pose graph of multiple robots, based on incremental smoothing and mapping, which guarantee that the goal is achieved with a higher convergence speed.
The authors of \cite{erol2016} use depth-cameras to perform SLAM and adopt a fusion technique, at a coordination center, that joins the point clouds generated by the sensors of multiple robots. The authors also highlight that task offloading, in the cloud, should take place to relieve onboard processing and power requirements.
Zhang et al. \cite{zhang2020} develop a cooperative approach to generate 3D global maps by fusing multiple 3D-lidar data sensors. Their framework is based on a local map construction and feature extraction, followed by common areas detection and global map merging. They further develop an accurate and low-complexity method to increase the accuracy of the resulting merged map.
The authors of \cite{li2024}, instead, devise a task allocation algorithm to perform cooperative mapping of an unexplored area with UAVs. They propose area segmentation in nonoverlapping portions, combined with a back-and-forth scanning pattern. The cooperative problem is then solved with the help of a genetic algorithm (GA). 
In \cite{xie2023} the authors propose a hierarchical multi-robot path planning pipeline, whose model involves i) a graph traversal problem solved by a GA, ii) a collision-free navigation problem solved with deep reinforcement learning, iii) pose estimation from lidar data, and iv) inter-robot data matching to close related graph loops.

\subsection{Radio-Aware Path Planning}

In \cite{debast2019}, coverage-aware algorithms have been proposed, to account for the maximization of the signal-to-interference-plus-noise ratio (SINR) along UAV trajectories. The authors used a lidar-scanned 3D map and 4G/5G base station (BS) locations, to simulate a realistic radio-coverage map of a city. Their optimal SINR-aware algorithm, with respect to a coverage-aware A* algorithm, has slightly higher SINR, but a much longer path length. 
Zhang \textit{et al.} \cite{zhang2021} build a SINR map that accounts for environment geometry and obstacles in the scenario, and also considers the traffic loading factor of each BS the UAVs are connected at. Then, they calculate a quantized grid of feasible path points where the SINR is over a fixed threshold, and run the Dijkstra algorithm to find the shortest path solution. 
In \cite{hu2022}, joint communication and sensing is used to shape the trajectory of UAVs, so that the ground environment is scanned with the highest possible precision using synthetic aperture radar techniques. The main target is to minimize flight energy consumption of the UAV, while keeping the focus on several landmarks of interest. This nonconvex problem is reformulated thanks to successive convex approximation and block coordinate descent into a series of feasibility checking subproblems.
Zhao \textit{et al.} \cite{zhao2021} develop a multi-UAV trajectory planning for energy-efficient coverage. The problem is posed as two multi-agent stochastic games, and it is then solved via a decentralized multi-UAV cooperative reinforcement learning technique. The UAVs are BSs in the sky, and ground users receive a guaranteed amount of information on their devices, provided that the energy consumption of UAVs is minimized, considering also a battery charging step.
Yang \textit{et al.} \cite{yang2019} modify the A* algorithm to include the effects of radio coverage while planning the trajectories of a UAV. Connectivity outage ratio and duration are considered among the additional costs to include in the path planning algorithm.
The trajectory and data exchange rate are maximized in \cite{you2020}, where the UAV has to fly next to sensors to collect data from them. Channel capacity is considered when deploying the trajectory, so that the data rate is kept above a threshold.

Recently, path planning strategies have included radio-based, handover-based, and offload-related side information. For example, Chai \textit{et al.} \cite{chai2025} propose an A* algorithm for 3-D UAV path planning, exploiting a static cellular infrastructure and precalculated radio maps. Jahandar \textit{et al.} \cite{jahandar2025} study the performance of handover mechanisms depending on mobility, network load, and service continuity in multi-access edge computing. Instead, Sato and Fujii \cite{sato2017} investigate radio-aware edge offloading, with load balancing and server selection prevailing over path planning.

With respect to the above-mentioned papers, our paper (i) embeds radio-related information directly into the costs of classical Dijkstra/A* path planning algorithms without employing connection feasibility constraints and without solving a continuous-path optimization problem \cite{debast2019,zhang2021,hu2022,yang2019,you2020}; (ii) enables a continuously-variable distance/connectivity trade-off without relying on SINR thresholds or access rate maximization \cite{debast2019,zhang2021,yang2019}; and (iii) adopts a deterministic planning framework, compatible with cooperative mapping and edge computing, without recurring to learning-based approaches \cite{zhao2021}.

Differently from detailed-radio-map-based planners such as \cite{chai2025}, our search strategy does not depend on the radio map construction method. Differently from handover-based or offloading-related approaches \cite{jahandar2025,sato2017}, our approach only requires a normalized scalar radio weight function, without using handover or offloading policies.

\section{Task Offloading in Edge Computing} \label{sec:taskoffload}

The considered edge computing scenario is amenable to perform task offloading from the AVs to the edge servers, namely, to reduce the latency of computationally heavy computer vision (CV) algorithms and to save energy for extended endurance.

In this scenario, a \textit{mission} for each AV is created on-line as the AV explores the environment with on-board sensors (cameras, lidars, radars, etc.). AVs may exploit multiple network interfaces with different radio access technologies (RAT) to communicate, such as for example IEEE 802.11ad for directional, high-speed, short-range, very-low-latency communication in the \SI{60}{\giga\hertz} band \cite{rappaport2017overview}, and 5G New Radio (NR), for omnidirectional, medium-to-high speed, long-range, low-latency communication in the sub-\SI{6}{\giga\hertz} band \cite{sun2016}. Due to the different adopted RATs, radio beams geometry may preclude the existence of a wireless link, even in case of direct visibility (i.e., no obstructions).

The AVs are equipped with embedded computing power, but in some cases it might be better to offload some of the tasks (such as object detection, SLAM, etc.) to external services \cite{detti-micro} instantiated in computational nodes (CN) or data centers (DC) \cite{ali2023,singh2023} by exploiting the wireless link offered by access points (AP) or BSs \cite{baruffa2024}.
Due to the strict communication latency requirements, edge computing is a potential solution to the problem of large network delays \cite{brehon2022}.

\begin{figure}
\centering
\includegraphics[width=0.9\columnwidth]{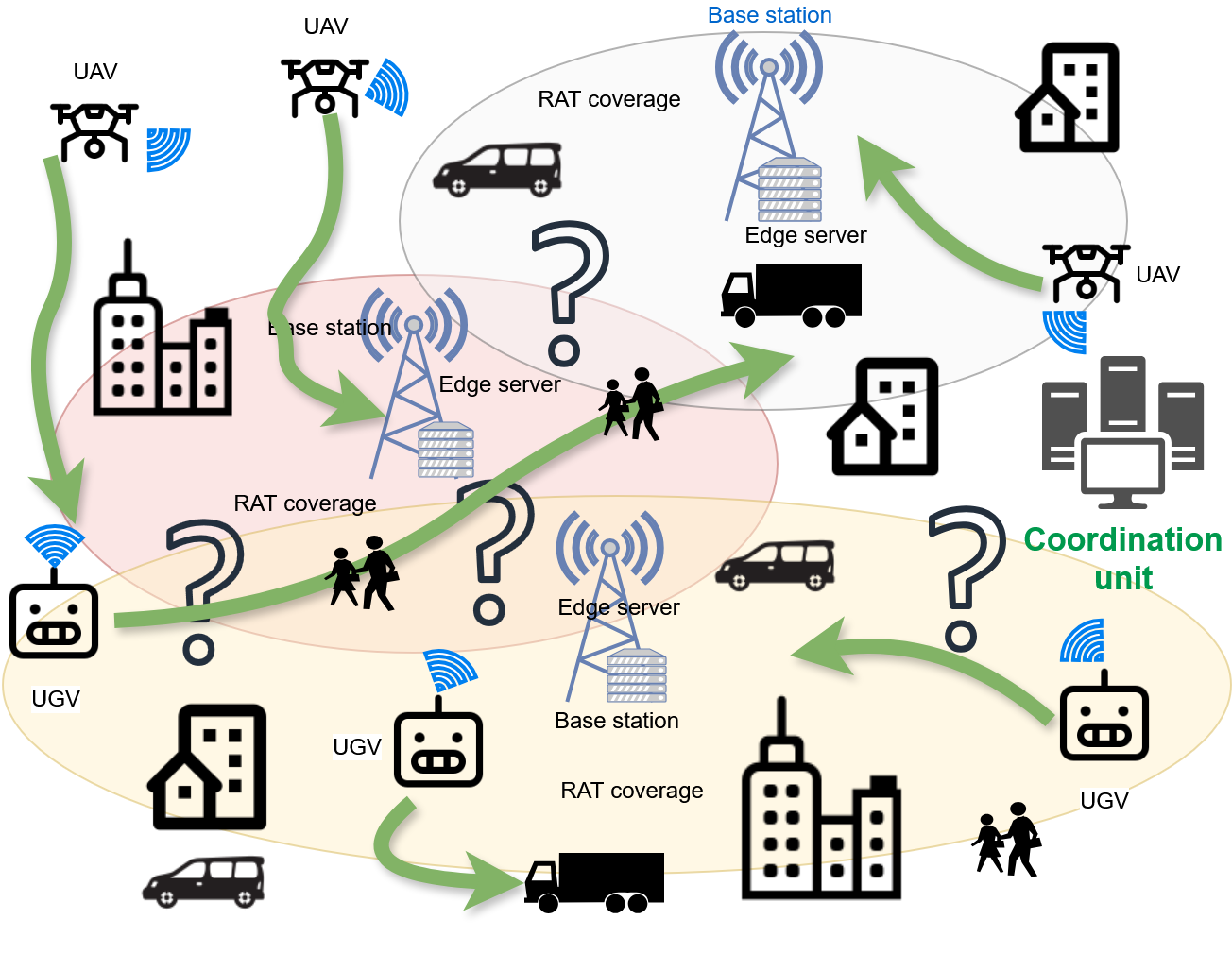}
\caption{Overview of the edge computing scenario.\label{fig:systemmodelwp4}}
\end{figure}

A concise representation of our task offloading model is given in Fig.~\ref{fig:systemmodelwp4}. The scenario is populated by a fleet of UAVs and unmanned ground vehicles (UGVs), which pursue a navigation or exploration mission in an environment populated with objects and obstacles, trying to map the environment \cite{yuan2020}. 

\section{Cooperative Mapping and Navigation} \label{sec:coopmap}

The exploration of a specified unknown area, or region, from AVs is a well-known duty in the scientific literature \cite{kumar2023}. We assume that the AVs, whose number is $N_\text{AV}$, are equipped with 2D range-based sensors \cite{holzhuter2023}, e.g., lidars or depth cameras, which cover a maximum range $\rho_\text{max}$.
In our sensing model, we consider measurement errors in the order of a few centimeters in range and a few tenths of a degree in angular resolution \cite{holzhuter2023}. 
Pose estimate errors, instead, are not considered, so as to extract the raw performance of the cooperative mapping and navigation algorithms. Such pose errors, generated for example by GPS or DGPS devices, may range from a few centimeters to a few meters, in outdoor environments \cite{ohno2004}.
In our scenario, the AVs start from positions placed on the boundary of the region to be explored.

The region of interest is represented by a discrete, ground-truth obstacle map matrix $\mathbf{B}$, of size $N\times N$: we use  $[\mathbf{B}]_\mathbf{n}$, where $\mathbf{n}=(n_1,n_2)$, as a shorthand to identify the entry of the matrix $[\mathbf{B}]$ that corresponds to a 2D geometric coordinate $(x,y)= (\Delta(n_2-1),\Delta(n_1-1))$, where $\Delta$ is the spatial precision. The value of $[\mathbf{B}]_\mathbf{n}$ is equal to $1$ when there is an obstacle in $(n_1,n_2)$, and it is equal to $0$ when $(n_1,n_2)$ is free from obstacles. We denote with $\hat{\mathbf{B}}(t)$ the obstacle map cooperatively estimated by the AVs at time $t$: when a point $(n_1,n_2)$ is still undecided, then the estimate is $[\hat{\mathbf{B}}(t)]_\mathbf{n} = 1/2$. Therefore, while the ground truth obstacle map $\mathbf{B}$ is binary, the estimated map $\hat{\mathbf{B}}(t)$ is ternary.

Each AV stores its own local copy $\hat{\mathbf{B}}_i(t)$ of the cooperatively estimated  obstacle map, which is periodically synchronized (with period $T_\text{S}$) with the main obstacle map $\hat{\mathbf{B}}(t)$, i.e., $\hat{\mathbf{B}}_i(t_{i,s}) \triangleq \hat{\mathbf{B}}(t_{i,s})$, where $t_{i,s}$, $s=1,2,\ldots$, are the synchronization time instants employed by the $i$th AV.
The main obstacle map $\hat{\mathbf{B}}(t)$ is updated, by fusion, when each $i$th AV sends its local copy $\hat{\mathbf{B}}_i(t)$, as
\begin{equation}
[\hat{\mathbf{B}}(t_{i,s} + \Delta t)]_\mathbf{n} = \begin{cases}
    [\hat{\mathbf{B}}_i(t_{i,s})]_\mathbf{n}, & [\hat{\mathbf{B}}_i(t_{i,s})]_\mathbf{n} \in \{0,1\}, \\
    [\hat{\mathbf{B}}(t_{i,s})]_\mathbf{n}, & \text{otherwise},
\end{cases}\label{eq:fusionrule}
\end{equation}
where $\Delta t$ represents the update lag.
The fusion rule in \eqref{eq:fusionrule} implements a last-observation strategy, with new measurements that overwrite previous values. This results in recent observations more representative than older ones, which is appropriate when there are mobile obstacles. Other strategies, such as majority voting or probabilistic occupancy, could also be employed \cite{howard2002}, at the cost of some increased complexity. 
We highlight that when two agents try to concurrently update the main map with their local copies, resource contention is applied to preserve map coherency.

The navigation service adds exploration waypoints in a serial fashion, one after another; specifically, once the current waypoint is reached, the AV moves through the known portion of the map to classify the still unknown positions.
Specifically, it chooses a new \textit{tentative} waypoint $\mathbf{w}^{\text{(new)}}$ as the nearest point lying\footnote{We choose the nearest undecided point since, with very large or  unlimited maps, it could be difficult to store the entire map in devices with limited resources.} in the yet-undecided area, with a small random offset, as
\begin{equation}
\mathbf{w}^{\text{(new)}} = \argmin_{\mathbf{n}: \, [\hat{\mathbf{B}}_i(t)]_\mathbf{n} = 1/2} \{ \Vert \mathbf{n} - \mathbf{p}(t) \Vert_2 \} + \boldsymbol{\xi},
\end{equation}
where $\mathbf{p}(t)$ is the current position of the AV in the map, and $\boldsymbol{\xi} = (\xi_1,\xi_2)$ is a random integer offset uniformly distributed in $[-5,5]$: this spatial randomization is applied to minimize the probability that two AVs choose the same waypoint as destination.
Moreover, a simple obstacle-avoidance strategy prevents each AV from colliding with obstacles in unexplored areas, and accounts for the vehicle dimensions, which is represented by a disc of radius $3$ pixels.
Since all AVs periodically receive the global obstacle map to update their own internal obstacle map, cooperation is intended here as the sharing of the same common obstacle map to decide new areas to explore. The obstacle map discovery process terminates when the undecided area is zero or when the AVs are blocked by insurmountable barriers to reach the yet undecided area.
The estimated obstacle map $\hat{\mathbf{B}}(t)$, which is the same for all users, is considered the output of the cooperative mapping algorithm at time $t$.

\subsection{Improved Obstacle Map}

Due to ranging issues and AV trajectories, the obstacle map could not correctly represent the position of an obstacle. Thus, we create a safe border around the detected obstacles by post-processing $\hat{\mathbf{B}}$.
First, the original obstacle map $\hat{\mathbf{B}}$, considered as an image, is lowpass filtered by a 2D filter $\mathbf{h}$. Then, the filtered obstacle map is optionally downsampled by $L \geq 1$, if complexity reduction is necessary. Finally, the resized obstacle map is binarized against a threshold $\tau$:\footnote{Value of $\tau$ can be chosen either empirically or using Otsu method \cite{liu2009}.} this obstacle map constitutes the actual obstacle map $\mathbf{O}$ to be used in the path planning phase. Mathematically,
\begin{equation}
[\mathbf{O}]_\mathbf{n} = \begin{cases}
1, & \left( \left( [\hat{\mathbf{B}}]_\mathbf{n} \circledast [\mathbf{h}]_\mathbf{n} \right) \downarrow L \right) > \tau,\\
0, & \left( \left( [\hat{\mathbf{B}}]_\mathbf{n} \circledast [\mathbf{h}]_\mathbf{n} \right) \downarrow L \right) \leq \tau ,
\end{cases}
\end{equation}
where $\circledast$ is the discrete 2D convolution operator on the $N \times N$ support, $\mathbf{h}$ is the 2D filter impulse response, and $\downarrow L$ is the uniform 2D downsampling operator that drops $L-1$ values out of $L$, for each variable $n_1$ and $n_2$.

\subsection{Performance Metrics}

Performance metrics deal with the amount of obstacle map coverage and the accuracy of obstacle map classification.
The obstacle coverage metric $C(t)$ represents the fraction of obstacle map that has been classified as free or occupied at time $t$, while its complement $\bar{C}(t)$ is the fraction of obstacle map that has not been explored yet. The obstacle coverage is defined as
\begin{equation}
C(t) = \frac{1}{N^2} \textstyle\sum_\mathbf{n} ([\hat{\mathbf{B}}(t)]_\mathbf{n} \neq \frac{1}{2}),
\end{equation}
where the sum extends over $\mathbf{n}=\{1,\ldots,N\} \times \{1,\ldots,N\}$, and the complement obstacle coverage $\bar{C}(t)$ is 
\begin{equation}
\bar{C}(t) = \frac{1}{N^2} \textstyle\sum_\mathbf{n} ([\hat{\mathbf{B}}(t)]_\mathbf{n} = \frac{1}{2}) = 1 - C(t).
\end{equation}
Additionally, we define the convergence time $T_\epsilon$ of the mapping algorithms as the time after which the complement obstacle coverage falls below $\epsilon$, that is
\begin{equation}
C(T_\varepsilon) \geq 1 - \epsilon. 
\end{equation} 

From $\hat{\mathbf{B}}(t)$ and $\mathbf{B}$ we can derive obstacle classification accuracy metrics. The \textit{false positive} obstacle map $\hat{\mathbf{B}}_{\text{FP}}(t)$ elements satisfy the relation
\begin{equation}
[\hat{\mathbf{B}}_{\text{FP}}(t)]_\mathbf{n} = ([\hat{\mathbf{B}}(t)]_\mathbf{n} = 1) \wedge ([\mathbf{B}]_\mathbf{n} = 0) ,
\end{equation}
which expresses the fact that a pixel is considered to be an obstacle, even if it has no obstacles in the real case.
For the \textit{false negative} obstacle map $\hat{\mathbf{B}}_{\text{FN}}(t)$, instead, 
\begin{equation}
[\hat{\mathbf{B}}_{\text{FN}}(t)]_\mathbf{n} = ([\hat{\mathbf{B}}(t)]_\mathbf{n} = 0) \wedge ([\mathbf{B}]_\mathbf{n} = 1) ,
\end{equation}
which expresses the fact that a pixel is considered to be free, even if it has an obstacle in the real case. 
Then, we can define the error rate in obstacle classification as
\begin{equation}
P_\text{e}(t) = \frac{1}{N^2} \textstyle\sum_\mathbf{n} ([\hat{\mathbf{B}}_{\text{FP}}(t)]_\mathbf{n} + [\hat{\mathbf{B}}_{\text{FN}}(t)]_\mathbf{n}).
\end{equation}

\section{Radio-Coverage-Aware Path Planning} \label{sec:pathplan}

We now consider the path planning problem.
The proposed algorithm plans the path of each AV independently of the others. The implemented navigation strategy avoids AV collisions when planned paths overlap.
Simple virtual collision sensors induce a temporary speed reduction and recovery strategy, after which the AVs resume their planned trajectories. Thus, inter-AV conflicts are not modeled in the path planner.
We assume that the global obstacle map (estimated during the cooperative mapping step) is available at the input of the AV path planner. To perform a radio-coverage-aware path planning, it is convenient to introduce two other types of maps. We define the \textit{path map} as the map that contains the planned path: this map, which is the output of the path planner, contains 1 in the pixels of the planned path and 0 in the pixels outside the planned trajectory. Since the path map is obtained as the concatenation of several segments, the path is characterized by several path endpoints between segments. Moreover, the radio coverage information is stored into another map, which we denote as \textit{radio weight map}: this map is used at the input of the path planner.

\begin{figure}[t!]
\centering
\includegraphics[width=0.98\columnwidth]{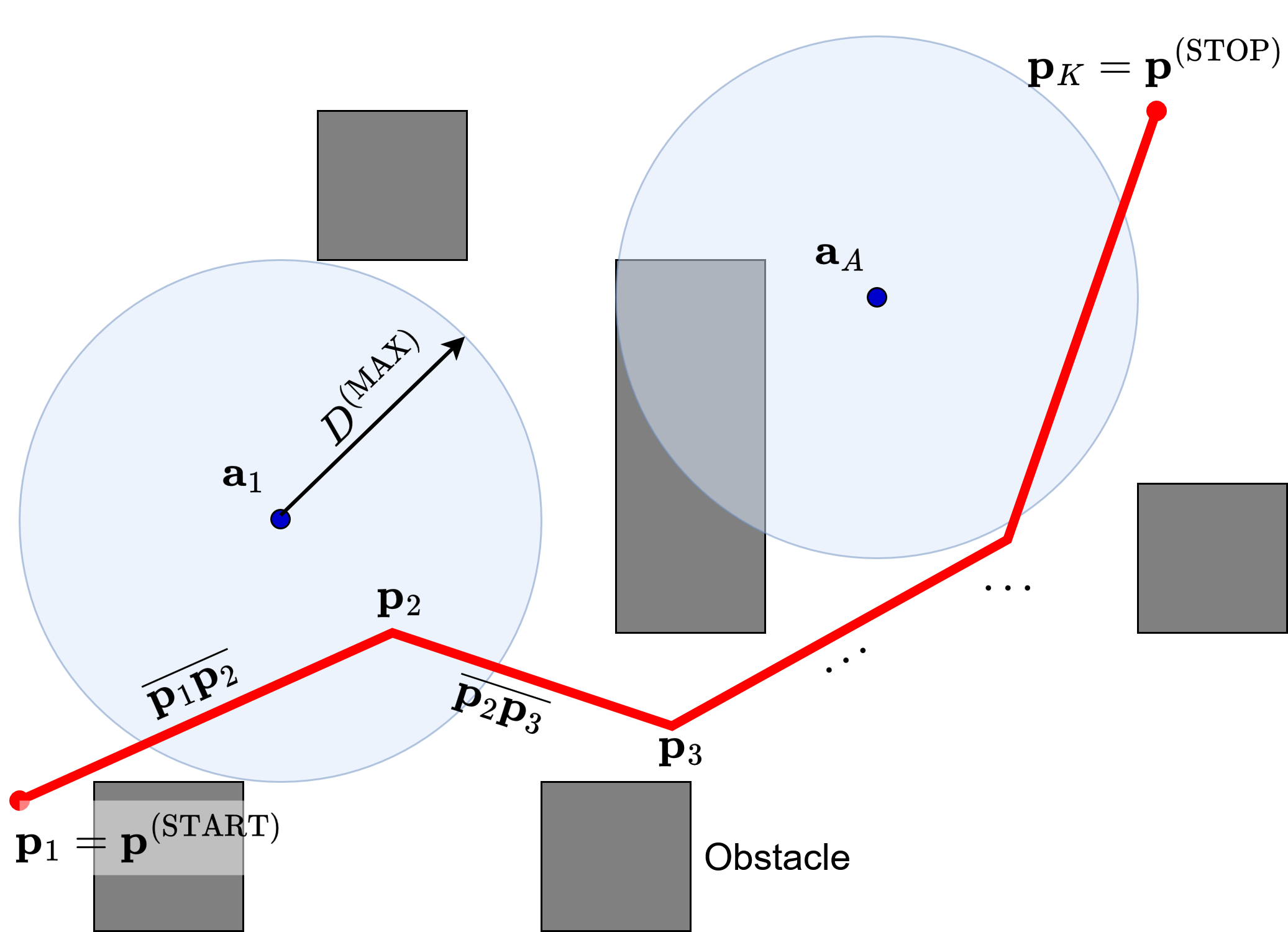}
\caption{Graphical overview of the radio-aware path planning model.\label{fig:model}}
\end{figure}

The system model used in the radio-coverage aware path planning problem is shown in Fig.~\ref{fig:model}. Given the start and stop points $\mathbf{n}=\mathbf{p}^{\textrm{(START)}}=\mathbf{p}_1$ and $\mathbf{n}=\mathbf{p}^{\textrm{(STOP)}}=\mathbf{p}_{K}$ in the obstacle map $\hat{\mathbf{B}}(T_\varepsilon) = \hat{\mathbf{B}}$, the problem is to plan a $K$-waypoint path, expressed by the segment path matrix $\mathbf{W}_K$ $= [\mathbf{p}_1^T \; \mathbf{p}_2^T \; \cdots \mathbf{p}_{K}^T ]^T$ with size $K \times 2$, that avoids the obstacles located in the obstacle map, i.e., the points where $[\hat{\mathbf{B}}]_\mathbf{n}>0$ (gray-colored areas). Additionally, the planned path shall minimize the traveled length, and shall maximize some radio-related metric related to the presence of $A$ APs, which are centered in $\mathbf{a}_1,\ldots,\mathbf{a}_{A}$ and are characterized by a radio coverage radius $D^{\textrm{(MAX)}}$. 

\begin{figure}
\centering
\includegraphics[width=\columnwidth]{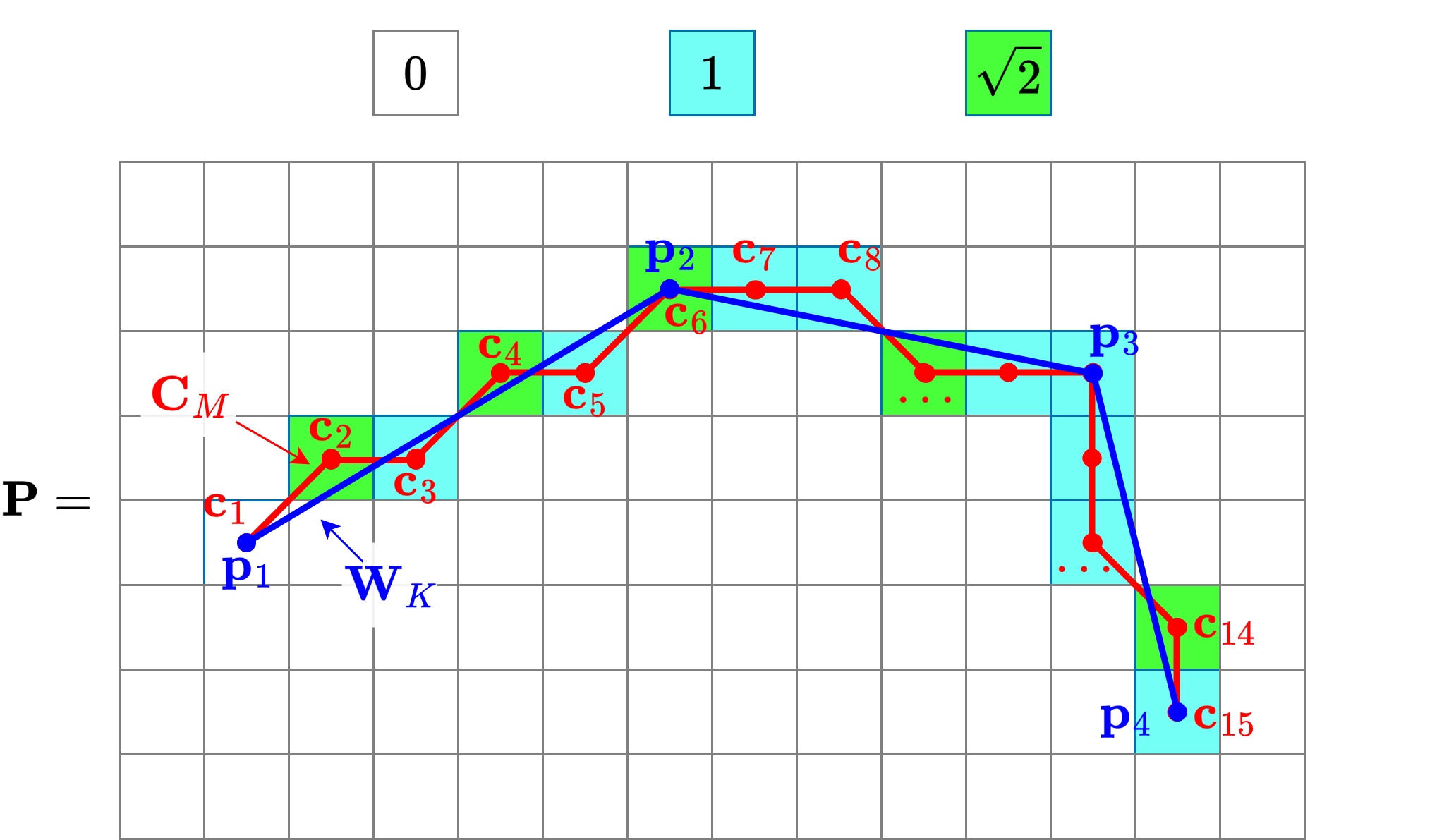}
\caption{Generation of the path $\mathbf{C}_M$ from path $\mathbf{W}_K$, and of the map $\mathbf{P}$ from path $\mathbf{C}_M$. Cyan and green pixels represent the synthesized path $\mathbf{C}_M$. \label{fig:segments}}
\end{figure}

To simplify the problem, it is convenient to transform the path $\mathbf{W}_K$, whose $K-1$ segments can be quite long, into a corresponding path $\mathbf{C}_M = [\mathbf{c}_1^T \; \mathbf{c}_2^T \; \cdots \mathbf{c}_{M}^T ]^T $, where the $M-1$ segments connect adjacent pixels only. Adjacent pixels are defined as those with $ \| \mathbf{c}_{m} - \mathbf{c}_{m-1} \|_\infty = 1$. The relation between the $\mathbf{W}_K$ and $\mathbf{C}_M$ paths is graphically shown in Figure~\ref{fig:segments}. The path $\mathbf{C}_M$ is obtained by discretizing all the path segments $\mathbf{W}_K$ with the Bresenham line drawing algorithm \cite{angel1991speeding}.
Finally, we define the $N\times N$ \textit{step distance matrix} $\mathbf{P}$, associated to the path with $\mathbf{C}_M$ endpoints, as
\begin{equation}
[\mathbf{P}(\mathbf{C}_{M})]_\mathbf{n} = \begin{cases}
0, & \mathbf{n} = \mathbf{c}_{1}, \\
\Vert \mathbf{c}_{m} - \mathbf{c}_{m-1} \Vert_2, & \mathbf{n} = \mathbf{c}_{m}, \;\; 2 \leq m \leq M, \\
0, & \mathbf{n} \neq \mathbf{c}_{m}, \;\; 1 \leq m \leq M .
\end{cases}\label{eq:PofPkappa}
\end{equation}
For instance, consider the path $\mathbf{W}_K$ with $K=4$ waypoints in Figure~\ref{fig:segments}, outlined in blue. The first path segment $[ \mathbf{p}_1^T \; \mathbf{p}_2^T]^T$ is expanded in the $6$-segment path $[\mathbf{c}_1^T \; \mathbf{c}_2^T \; \mathbf{c}_3^T \; \mathbf{c}_4^T \;\mathbf{c}_5^T \; \mathbf{c}_6^T ]^T$: the other segments $[ \mathbf{p}_2^T \; \mathbf{p}_3^T]^T$ and $[ \mathbf{p}_3^T \; \mathbf{p}_4^T]^T$ are expanded similarly. Then, the final path is $\mathbf{C}_M = [ \mathbf{c}_1^T \; \ldots \; \mathbf{c}_{15}^T]^T$, with $M=15$, represented in red color. The final step distance matrix $\mathbf{P}$ contains the value $\sqrt{2}$ in the green-colored points, the value $1$ in the cyan-colored points, while it is $0$-valued in all other blank points.

\subsection{Radio Coverage Weight Map}

The radio coverage weight map matrix $\mathbf{R}$ measures the effectiveness of the radio coverage on the considered regions.
Since every AV can connect to multiple APs, it can choose the one providing the best radio coverage, so the radio weight map depends on the number and placement of the APs as
\begin{equation}
[\mathbf{R}]_\mathbf{n} = \max_{j=1,\ldots,A} [\mathbf{R}^{(j)}]_\mathbf{n},
\end{equation}
where
\begin{equation}
[\mathbf{R}^{(j)}]_\mathbf{n} = \begin{cases}
[\tilde{\mathbf{R}}^{(j)}]_\mathbf{n}, & \Vert \mathbf{a}_{j} - \mathbf{n} \Vert_2 \leq D^{\textrm{(MAX)}}, \\
0, & \textrm{else} ,
\end{cases} \label{eq:costmapap}
\end{equation}
are the elements of the radio weight map generated by the $j$th AP located in $\mathbf{a}_{j} \in \{1,\ldots,N\}^2$, which gives null costs outside of the disc with radius $D^{\textrm{(MAX)}}$ pixels.
The AP radio weight $[\tilde{\mathbf{R}}^{(j)}]_\mathbf{n}$ can be defined considering different radio related aspects, and we assume they are normalized in a $[0,1]$-range metric.

The simplest possible metric is the \textit{on-off} radio weight map
\begin{equation}
[\tilde{\mathbf{R}}^{(j)}]_\mathbf{n} = 1 
\end{equation}
that, when inserted in \eqref{eq:costmapap}, indicates whether the point $\mathbf{n}$ is inside or outside of the AP coverage radius.
Differently, considering the AP-AV distance, the \textit{amplitude}-related radio weight map is 
\begin{equation}
[\tilde{\mathbf{R}}^{(j)}]_\mathbf{n} = \frac{1}{\Vert \mathbf{a}_{j} - \mathbf{n} \Vert_2^\gamma},
\end{equation}
which weights the distance from the AP in terms of a decaying received amplitude depending on the exponent $\gamma$. 
Then, we propose a \textit{capacity}-related radio weight map
\begin{equation}
[\tilde{\mathbf{R}}^{(j)}]_\mathbf{n} = 1 - \frac{\log_{2} \Vert \mathbf{a}_{j} - \mathbf{n} \Vert_2}{ \log_{2} D^{\textrm{(MAX)}}},
\end{equation}
which uses a Shannon-like informative capacity representation, normalized to vary between $0$ (at the disc border) and $1$ (near the disc center). 
Finally, the last metric uses a \textit{tent}-shaped decaying radio weight, defined as
\begin{equation}
[\tilde{\mathbf{R}}^{(j)}]_\mathbf{n} = \left( 1 - \frac{\Vert \mathbf{a}_{j} - \mathbf{n} \Vert_2}{ D^{\textrm{(MAX)}}} \right)^\beta,
\end{equation}
where $\beta<1$ represents the exponent of the tent function. 

The radio weight at the AP position $\mathbf{a}_{j}$ is set to zero to avoid singularities, also consistent with the fact that AP locations are generally not accessible.

The conceived radio weight surfaces are simple and normalized in the $[0,1]$ range, making them independent of propagation models and environments. More physically-based radio maps can be obtained by ray-tracing techniques or statistical channel models, so as to integrate blockage, interference, AP load, or handover effects (possibly left as future work).

\section{Radio-Aware Path Planning Problem} \label{sec:rapaplanprob}

We now formally define the optimization problem related to the coverage-aware path planning scenario.
Given the starting point $\mathbf{p}^{\textrm{(START)}}$ and the stopping point $\mathbf{p}^{\textrm{(STOP)}}$, the problem is to find the $M$ waypoints $\mathbf{C}_M$, so that the multi-segment path has minimum length, without overlapping the obstacles in $\mathbf{O}$, and to jointly maximize the average experienced radio coverage, as defined by the radio weight map $\mathbf{R}$.

This multi-objective integer programming (MOIP) problem is expressed by 
\begin{subequations}
\begin{align}
\min_{\mathbf{C}_M, M} \quad & \left[ F_1(\mathbf{C}_M), -F_2(\mathbf{C}_M) \right], \\
\textrm{s.t.} \quad & \mathbf{c}_m \in \{1,\ldots,N\}^2, \; \forall m,\\
\quad & \mathbf{c}_1 = \mathbf{p}^{\textrm{(START)}}, \\
\quad & \mathbf{c}_{M} = \mathbf{p}^{\textrm{(STOP)}}, \\
\quad & \| \mathbf{c}_{m} - \mathbf{c}_{m-1} \|_\infty = 1, \; \forall m, \label{eq:45degconstr} \\
\quad &  [\mathbf{O}]_\mathbf{n} \cdot [\mathbf{P}(\mathbf{C}_{M})]_\mathbf{n} = 0, \; \forall \mathbf{n}, \label{eq:obstconstr}
\end{align} \label{eq:mimp}
\end{subequations}
where  
\begin{align}
F_1(\mathbf{C}_M) &= \textstyle \sum_{m=2}^{M} \Vert \mathbf{c}_m - \mathbf{c}_{m-1} \Vert_2 \nonumber \\
&= \textstyle \sum_{\mathbf{n}} [\mathbf{P}(\mathbf{C}_{M})]_\mathbf{n},
\label{eq:F1}
\end{align}
with $\mathbf{n} \in \{1,\ldots,N\}^2$, is the total Euclidean length of the path $\mathbf{C}_M$,
\begin{align}
F_2(\mathbf{C}_M) &= \textstyle \sum_{m=2}^{M} [\mathbf{R}]_{\mathbf{c}_{m}} \Vert \mathbf{c}_m - \mathbf{c}_{m-1} \Vert_2 \nonumber \\
&= \textstyle \sum_\mathbf{n} [\mathbf{R}]_\mathbf{n} \cdot [\mathbf{P}(\mathbf{C}_{M})]_\mathbf{n}, \label{eq:F2}
\end{align}
is a metric representing the radio coverage weight experienced along the traveled path, \eqref{eq:45degconstr} constrains the movement to one pixel in the horizontal, vertical, or diagonal direction, and \eqref{eq:obstconstr} represents the obstacle avoidance constraint.

The performance of the path planner is evaluated using the optimal discrete path $\mathbf{C}_M^*$, from which we calculate the traveled distance $D = F_1(\mathbf{C}_M^*)$ and the traveled radio weight $R' = F_2(\mathbf{C}_M^*)$.

\section{Proposed Path Planning Solutions} \label{sec:propaplansol}

In order to solve this MOIP problem, the multi-objective Pareto-optimality concept \cite{hansen2013} is used to convert \eqref{eq:mimp} into a single-objective problem with the help of a $\lambda \in [0,1]$ parameter, as expressed by
\begin{equation}
\argmin_{\mathbf{C}_M} \; [\lambda F_1(\mathbf{C}_M) - (1-\lambda) F_2(\mathbf{C}_M)] . \label{eq:lambdaF1F2}
\end{equation}
This is equivalent to
\begin{equation}
\argmin_{\mathbf{C}_M} \; g(\mathbf{C}_M), \label{eq:moipbeta} 
\end{equation}
where
\begin{equation}
g(\mathbf{C}_M) = F_1(\mathbf{C}_M) -\alpha F_2(\mathbf{C}_M), \label{eq:costOD}
\end{equation}
and $\alpha=1/\lambda-1$, with $\alpha \in [0,+\infty)$, is a coefficient that favors either the traveled distance or the experienced radio weight in the optimal solution.
Note that $g(\mathbf{C}_M)$ is cumulative since, for \eqref{eq:F1}--\eqref{eq:F2},
\begin{align}
g(\mathbf{C}_M) & = \textstyle \sum_{m=2}^{M} \max \{ 1 - \alpha [\mathbf{R}]_{\mathbf{c}_{m}}, 0 \} \Vert \mathbf{c}_m - \mathbf{c}_{m-1} \Vert_2  \nonumber \\
& = \textstyle \sum_{\mathbf{n}} \max \{ 1 - \alpha [\mathbf{R}]_\mathbf{n} , 0 \} [\mathbf{P}(\mathbf{C}_{M})]_\mathbf{n}. \label{eq:gcumulative}
\end{align}
The $\max\{\}$ operator in \eqref{eq:gcumulative} prevents the cost function from becoming negative when $\alpha>1$ or, equivalently, $\lambda < 1/2$ in \eqref{eq:lambdaF1F2}, and lets the distance objective function $F_1$ dominate the radio objective function $F_2$.

\begin{figure}
\centering
\includegraphics[width=0.9\columnwidth]{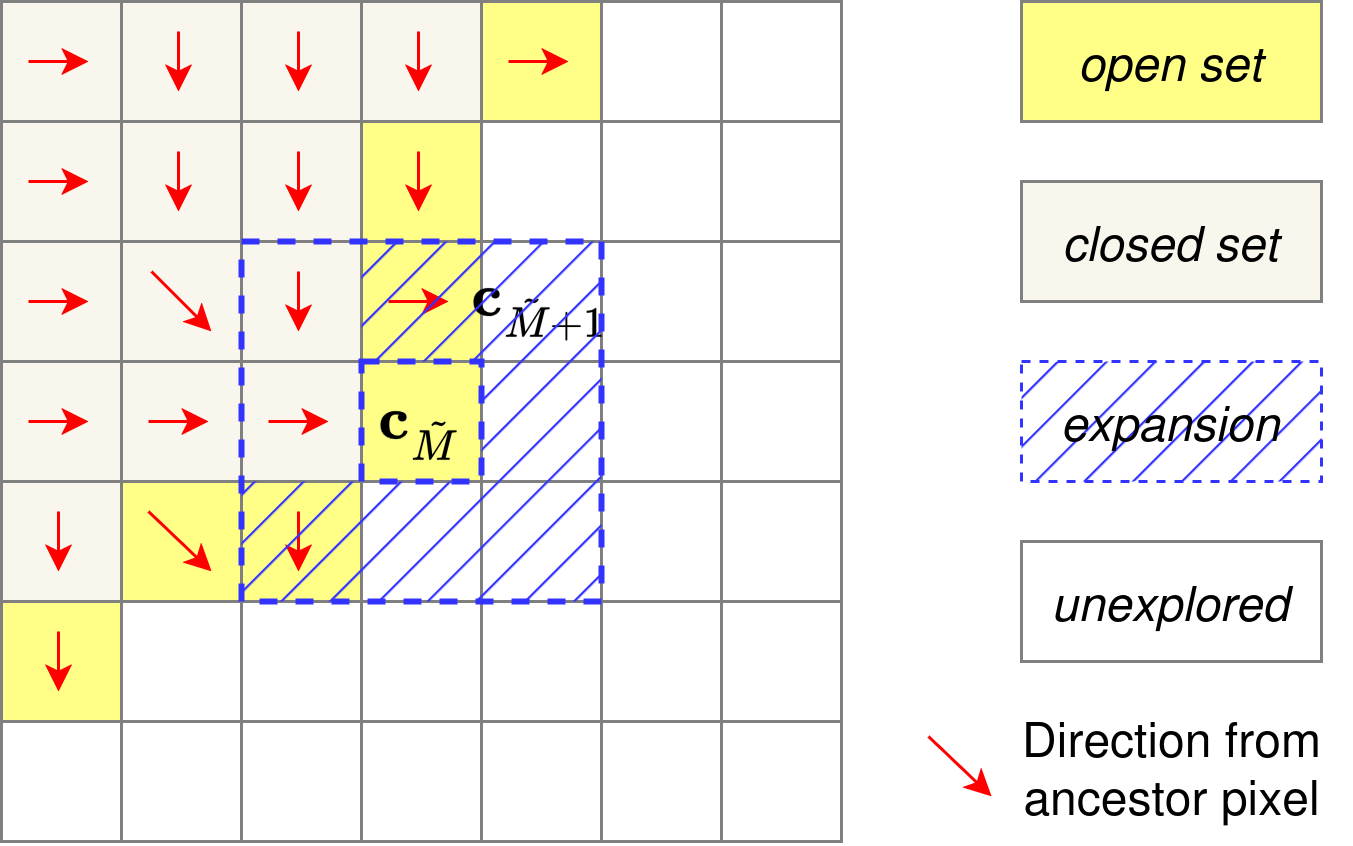}
\caption{Expansion of the frontier in the Dijkstra/A* algorithm.\label{fig:frontier}}
\end{figure}

Before considering the general cost function \eqref{eq:costOD}, we first consider the shortest-path problem (SPP), which is a special case of \eqref{eq:costOD} with $\alpha=0$. 
It is well known that the solution to the SPP is the Dijkstra algorithm \cite{ahuja1990}, which solves the minimization when the cost function is 
\begin{equation}
g^{\text{(OD)}}(\mathbf{C}_M) = F_1(\mathbf{C}_M).
\end{equation}
The original Dijkstra (OD) algorithm extends the \textit{frontier}, or \textit{open set}, with pixels chosen towards the ``outside'', and each pixel of the already explored set, the \textit{closed set}, is associated to a complete and unique route to the starting point, which optimizes a prescribed cost function (Fig.~\ref{fig:frontier}).
Specifically, given a tentative point $\mathbf{c}_{\tilde{M}}$ of the path on the current frontier, and $\mathbf{c}_{\tilde{M}+1}$ on the new expanded frontier, the OD algorithm uses the Euclidean cost

\begin{align}
g^{\text{(OD)}}(\mathbf{C}_{\tilde{M}+1}) =& g^{\text{(OD)}}(\mathbf{C}_{\tilde{M}}) + \Vert \mathbf{c}_{\tilde{M}+1} - \mathbf{c}_{\tilde{M}} \Vert_2 \label{eq:oldf}
\end{align}
to decide which is the optimum path ending in a point on the frontier. The expansion and cost calculation is performed for all the points on the current frontier; then, each point on the new expanded frontier is associated to the path with minimum distance which leads to the point.
We now propose two new OD-based algorithms, with different computational complexity, that are capable to solve the general problem expressed by \eqref{eq:moipbeta}--\eqref{eq:costOD} with $\alpha\neq 0$.

\subsection{Weighted Dijkstra Algorithm}

Similarly to the previous subsection, we denote with $\mathbf{c}_{\tilde{M}+1}$ a tentative point on the expanded frontier and, by using \eqref{eq:gcumulative}, we obtain the weighted Dijkstra (WD) algorithm expressed by
\begin{align}
g^{\text{(WD)}}(\mathbf{C}_{\tilde{M}+1}) &= g^{\text{(WD)}}(\mathbf{C}_{\tilde{M}}) \label{eq:costwithradioapprox} \\
&+ \max\{ 1 - \alpha [\mathbf{R}]_{\mathbf{c}_{\tilde{M}+1}} , 0\} \cdot \Vert \mathbf{c}_{\tilde{M}+1} - \mathbf{c}_{\tilde{M}} \Vert_2 . \nonumber
\end{align}
Note that \eqref{eq:costwithradioapprox} reduces the Euclidean cost $\Vert \mathbf{c}_{\tilde{M}+1} - \mathbf{c}_{\tilde{M}} \Vert_2$ in \eqref{eq:oldf} by a factor $\max \{ 1- \alpha [\mathbf{R}]_{\mathbf{c}_{\tilde{M}+1}} , 0 \}$. Therefore, points with larger coverage weight $[\mathbf{R}]_{\mathbf{c}_{\tilde{M}+1}}$ produce a stronger reduction of the \textit{algorithmic} distance, thus promoting the directions with higher radio weights.

\subsection{Weighted A* Algorithm}

The A* algorithm \cite{foead2021systematic} is a simplification of the Dijkstra procedure, conceived to reduce its computational complexity. While Dijkstra systematically explores all points on the frontier to expand the closed set, A* only explores the neighborhood of a convenient point $\mathbf{c}_{\tilde{M}}$ on the frontier, selected by minimizing the function
\begin{equation}
f(\mathbf{C}_{\tilde{M}},\mathbf{p}^{\textrm{(STOP)}}) = g(\mathbf{C}_{\tilde{M}}) + \gamma(\mathbf{c}_{\tilde{M}}, \mathbf{p}^{\textrm{(STOP)}}) , \label{eq:fofastar}
\end{equation}
where $\gamma(\mathbf{c}_{\tilde{M}},\mathbf{p}^{\textrm{(STOP)}})$ is an \textit{admissible heuristic}, i.e., a lower bound of the total cost to reach the ending point 
$\mathbf{p}^{\textrm{(STOP)}}$. Intuitively, $\gamma(\mathbf{c}_{\tilde{M}},\mathbf{p}^{\textrm{(STOP)}})$ steers the expansion of the frontier towards a preferred direction. In the original A* (OA) algorithm, the heuristic is
\begin{equation}
\gamma(\mathbf{c}_{\tilde{M}},\mathbf{p}^{\textrm{(STOP)}}) = \| \mathbf{p}^{\textrm{(STOP)}} - \mathbf{c}_{\tilde{M}} \|_2 ,   \label{eq:gammaoa}
\end{equation}
the Euclidean distance to the ending point. However, to take radio costs into account, the function $g(\mathbf{C}_{\tilde{M}})$ in \eqref{eq:fofastar} should consider not only the distance, but also the radio weights. Thus, to steer the radio-aware expansion of the frontier towards a preferred direction, we replace the OA heuristic in \eqref{eq:gammaoa} with the new heuristic
\begin{equation}
\gamma^\text{(WA)}(\mathbf{c}_{\tilde{M}},\mathbf{p}^{\textrm{(STOP)}}) = \max \{ 1- \alpha , 0 \} \cdot \| \mathbf{p}^{\textrm{(STOP)}} - \mathbf{c}_{\tilde{M}} \|_2, \label{eq:gammawa}
\end{equation}
where WA stands for weighted A* algorithm.
Indeed, since $[\mathbf{R}]_\mathbf{n} \in [0,1]$, we have  $\max \{ 1 - \alpha , 0 \} \leq \max \{ 1 - \alpha [\mathbf{R}]_\mathbf{n} , 0 \}$. For any feasible path from $\mathbf{c}_{\tilde{M}}$ to $\mathbf{p}^{\textrm{(STOP)}}$, and by the triangle inequality,
\begin{align}
 \textstyle \sum_{m = \tilde{M}+1}^{M}  & \max \{ 1 - \alpha [\mathbf{R}]_{\mathbf{c}_{m}}, 0 \}  \Vert \mathbf{c}_m - \mathbf{c}_{m-1} \Vert_2 \nonumber \\
 & \geq  \max \{ 1- \alpha , 0 \} \| \mathbf{p}^{\textrm{(STOP)}} - \mathbf{c}_{\tilde{M}} \|_2 \, .
\end{align}
Therefore, \eqref{eq:gammawa} is a lower bound of the remaining cost and is admissible. Using similar arguments, the heuristic is shown to be monotone (consistent). Thus, WA remains optimal when using the cost function in \eqref{eq:gcumulative}. 

\subsection{Weighted Suboptimal Algorithm}

While \eqref{eq:gammawa} guarantees admissibility and consistency, it is inherently conservative and may reduce steering capability for large values of $\alpha$. A more aggressive radio-aware frontier expansion can be achieved through the weighted suboptimal (WS) heuristic
\begin{align}
\gamma^\text{(WS)} & (\mathbf{c}_{\tilde{M}},\mathbf{p}^{\textrm{(STOP)}}) \nonumber \\
& = \max \{ 1- \alpha [\mathbf{R}]_{\mathbf{c}_{\tilde{M}}} , 0 \} \cdot \| \mathbf{p}^{\textrm{(STOP)}} - \mathbf{c}_{\tilde{M}} \|_2 . \label{eq:gammaws}
\end{align}
Unlike \eqref{eq:gammawa}, the heuristic \eqref{eq:gammaws} includes the radio weight and is not generally admissible, since the radio coverage distribution along the remaining path may differ from that observed at $\mathbf{c}_{\tilde{M}}$. Consequently, optimality with respect to \eqref{eq:gcumulative} cannot be guaranteed. 

Nevertheless, the WS heuristic more strongly favors well-covered radio regions: in practice, it reduces the number of explored points and increases the radio experience. Therefore, the WS heuristic \eqref{eq:gammaws} represents another complexity-optimality trade-off, in addition to the WA heuristic \eqref{eq:gammawa}.

\section{Simulation Results} \label{sec:simresults}

We calculate the performance of the proposed obstacle mapping and path planning algorithms by simulation in a simple scenario, characterized by a variable number of AVs and a fixed number of obstacles. First, we illustrate simulation results on cooperative mapping, to acquire the obstacle map successively exploited during path planning. Then, we focus on the simulated performance of the path planning algorithms proposed in this paper, in comparison with existing path planning algorithms. 

In the following simulations, the AV pose is assumed to be perfectly known, to evaluate the performance of the proposed cooperative mapping and radio-aware path planning algorithms. Nevertheless, our setup can accept pose estimates affected by typical GPS or DGPS localization errors. Similarly, while ideal range measurements are used in the simulations, the sensing model can account for range and angular errors.

\subsection{Cooperative Mapping Results}\label{sec:coopmapres}

\begin{figure}
\centering
\includegraphics[width=\columnwidth]{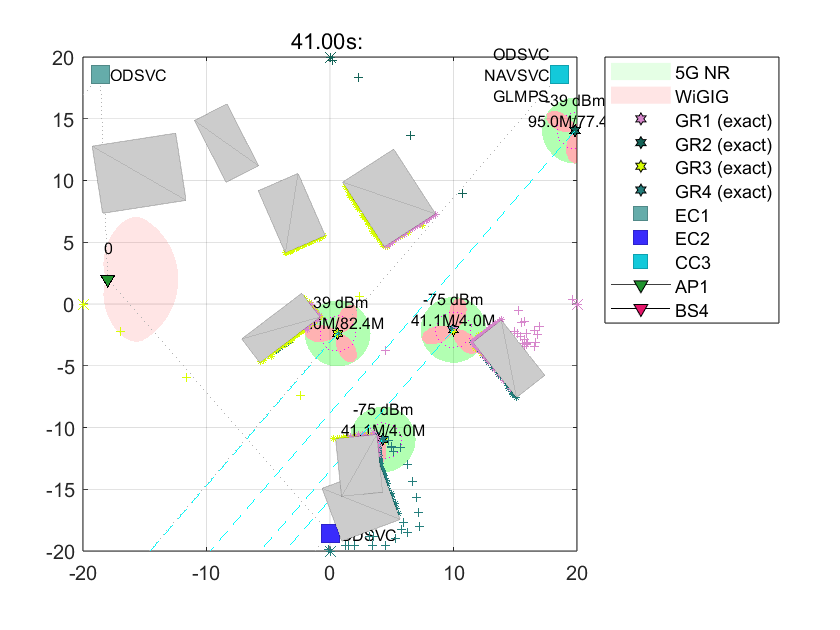}
\caption{Localization of the AVs and BSs in the obstacle map, with obstacles and lidar intercept points.\label{fig:uvs2dview}}
\end{figure}

Figure~\ref{fig:uvs2dview} shows the simulated system at a glance. It is composed of $4$ AVs (ground robots GR1, GR2, GR3, and GR4), a WiGig AP, a 5G NR BS, and three CNs (EC1 and EC2 are on the edge, while CC3 is in the cloud). Each AV starts its exploration task from a different side of the obstacle map (north, east, south, west) and uses a lidar range sensor with a range $\rho_\text{max}=\SI{12}{\meter}$. The adopted obstacle map, which has size of $40\times40$ m, is represented with a precision $\Delta=\SI{0.1}{\meter}$, to obtain $N=400$ pixels per dimension. There are $7$ box-shaped obstacles in the area to explore. Each AV synchronizes its local obstacle map with a coordination service placed in the cloud node CC3, which also hosts the centralized navigation service. All CNs host the task offloading service, which is devoted to object detection using some CV algorithm.  The starting times of the obstacle map synchronization are randomized, i.e., $t_{i,1} \sim \mathcal{U}(0,T_\text{S})$, with synchronization period $T_\text{S}=\SI{3}{\second}$: the simulation time advances in discrete time steps of $\Delta t = 0.25$s. The cooperative mapping is stopped when the missing coverage falls below $\varepsilon=\SI{1e-4}{}$, i.e., when $99.99 \%$ of the area has been explored.

\begin{figure}[t!]
\centering
\subfloat[AV1]{\label{fig:uv1t0}\includegraphics[width=0.24\columnwidth]{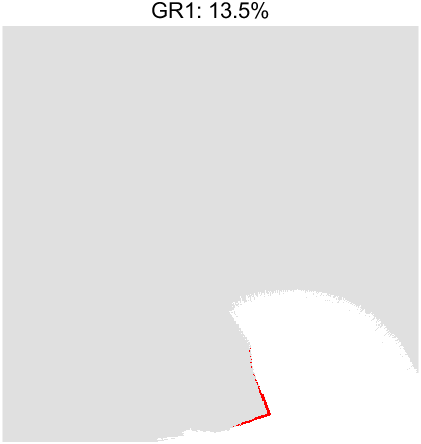}}
\hfill
\subfloat[AV2]{\label{fig:uv2t0}\includegraphics[width=0.24\columnwidth]{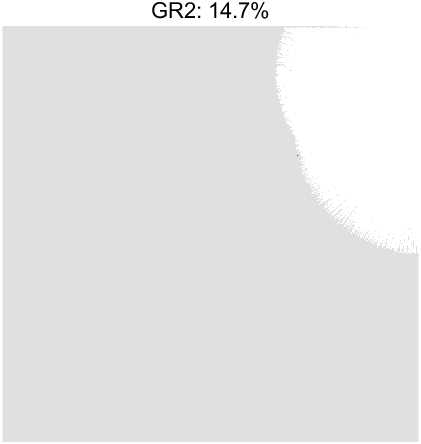}}
\hfill
\subfloat[AV3]{\label{fig:uv3t0}\includegraphics[width=0.24\columnwidth]{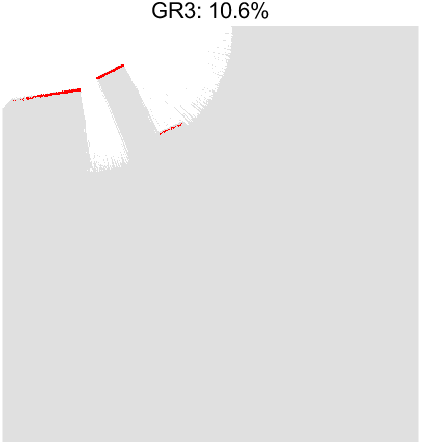}}
\hfill
\subfloat[AV4]{\label{fig:uv4t0}\includegraphics[width=0.24\columnwidth]{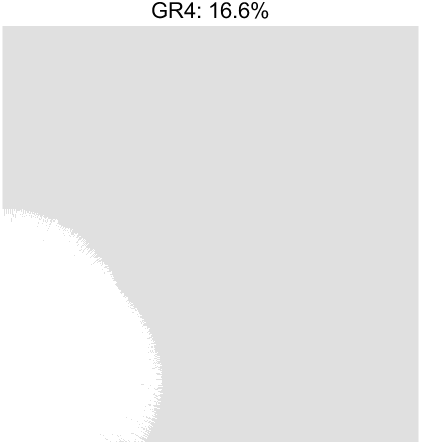}}\\
\subfloat[AV1]{\includegraphics[width=0.24\columnwidth]{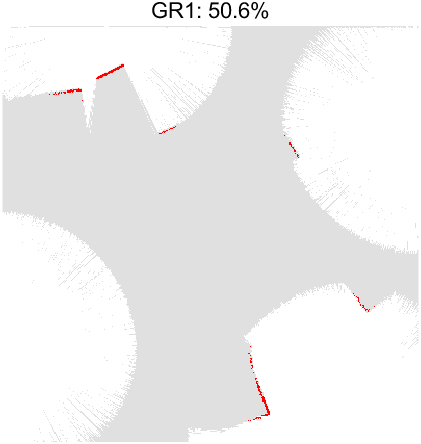}\label{fig:uv1t1}}
\hfill
\subfloat[AV2]{\includegraphics[width=0.24\columnwidth]{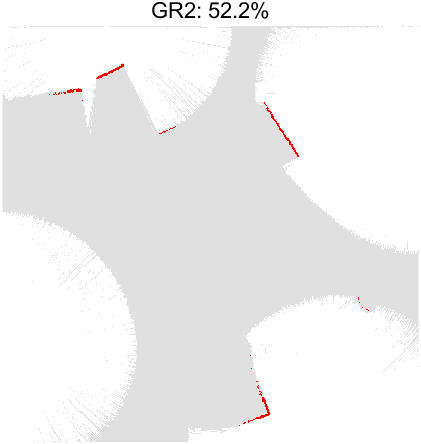}\label{fig:uv2t1}}
\hfill
\subfloat[AV3]{\includegraphics[width=0.24\columnwidth]{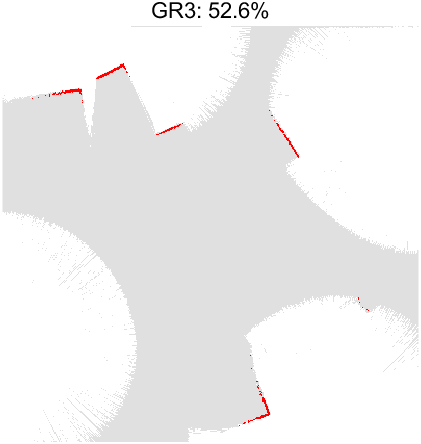}\label{fig:uv3t1}}
\hfill
\subfloat[AV4]{\includegraphics[width=0.24\columnwidth]{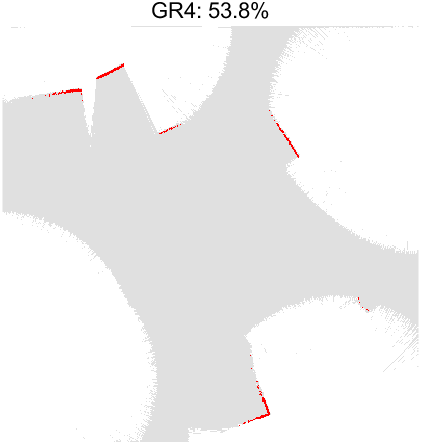}\label{fig:uv4t1}}
\caption{Local obstacle maps $\hat{\mathbf{B}}_i(t)$ for the AVs, before (a, b, c, d) and after (e, f, g, h) the first synchronization. Obstacles are colored in red for clarity.}
\label{fig:uvst1}
\end{figure}

Figures~\ref{fig:uv1t0}--\ref{fig:uv4t0} show the local obstacle maps $\hat{\mathbf{B}}_i(t)$ of the $4$ AVs, immediately before the first synchronization instant $t_{i,1}$. Each figure reports also the coverage $C(t)$ of the local obstacle maps, expressed as percentages. For improved clarity in the figures, detected obstacles are colored in red, explored areas in white, and unexplored regions in gray. Figures~\ref{fig:uv1t1}--\ref{fig:uv4t1}, instead, show the local obstacle maps obtained after the fusion of the maps of all AVs, when the first synchronization event has occurred.

\begin{figure}
\centering
\includegraphics[width=0.6\columnwidth]{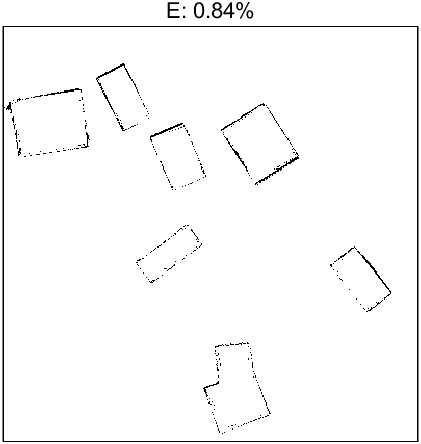}
\caption{Error map for scenario in Figure~\ref{fig:uvs2dview}.\label{fig:perrmap8888}}
\end{figure}

At the end of the exploration ($t = T_\epsilon$), Figure~\ref{fig:perrmap8888} shows the error pixels, colored in black for improved visibility. The errors are located around the obstacle boundaries, due to misclassified pixels. These boundary errors outline the safe navigation area, with limited impact on path layouts. The final obstacle mapping error probability is $P_\text{e}(T_\epsilon)=0.84\%$.

\begin{figure}
\centering
\includegraphics[width=\columnwidth]{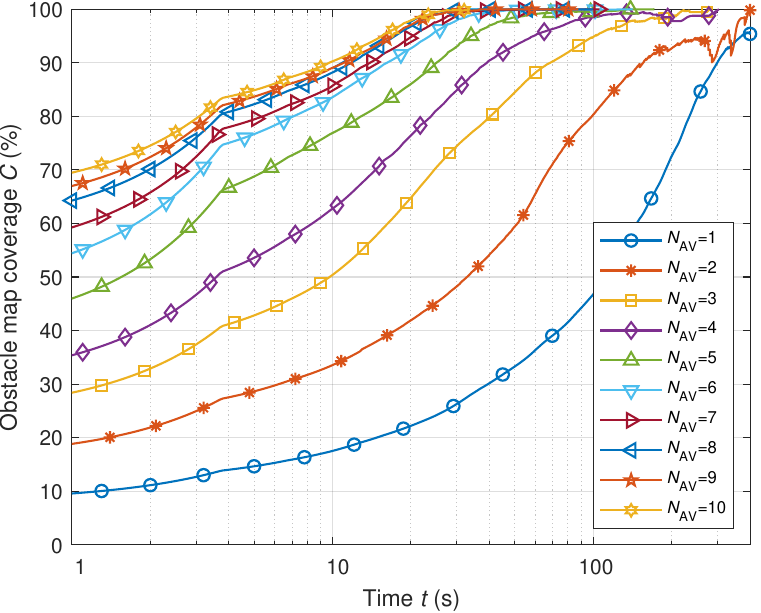}
\caption{Obstacle map coverage as a function of the number of AVs.\label{fig:coverageavg}}
\end{figure}

\begin{figure}
\centering
\includegraphics[width=\columnwidth]{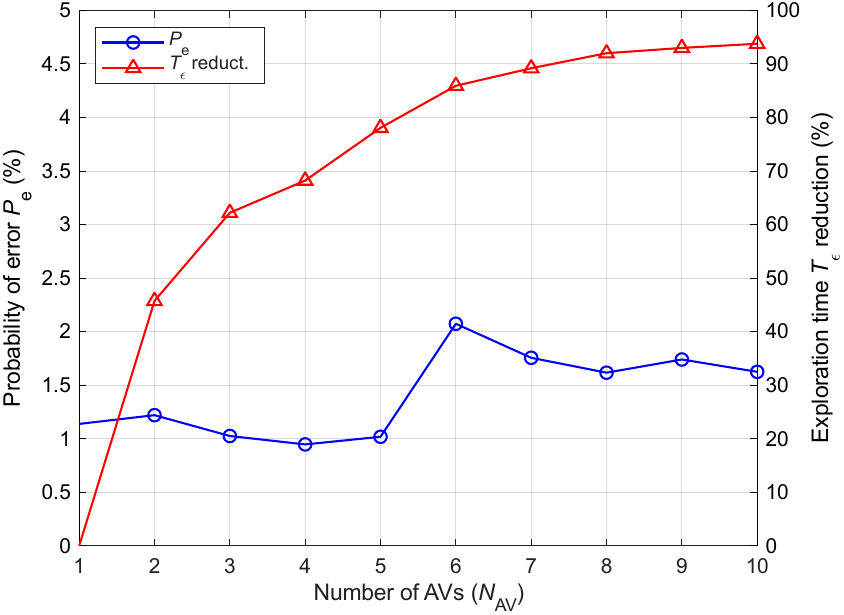}
\caption{Exploration time and probability of obstacle map error versus the number of AVs.\label{fig:perroravg}}
\end{figure}

Upon varying the number of AVs, and keeping fixed all the other parameters, we obtain the coverage as a function of $N_\text{AV}$, shown in Figure~\ref{fig:coverageavg}. For the obstacle map under observation, there is a significant improvement in obstacle mapping times until $N_\text{AV}=6$, while for higher values there is only a slight gain. 
Figure~\ref{fig:perroravg} shows the probability of obstacle map error versus the number of AVs: the probability of error is roughly between $1\%$ and $2\%$. This error probability is considered sufficiently low for our path planning purposes: indeed, the large majority of erroneous pixels are close to the borders of the obstacles, but this does not produce path planning problems, because we included in the planning algorithms a safety zone, i.e., the AV trajectories have to respect a minimum distance, greater than zero, from the border of the obstacles.

Figure~\ref{fig:perroravg} also shows the reduction of the exploration time $T_\epsilon$ versus the number of AVs. With respect to $N_\text{AV}=1$, the exploration time is reduced by $45\%$ with $2$ AVs, and by $90\%$ when $N_\text{AV}=7$: adding more AVs does not yield a significant reduction of the exploration time.

Figures~\ref{fig:coverageavg}~and~\ref{fig:perroravg} testify that cooperation among AVs significantly reduces the exploration time, thereby producing obstacle maps (to be used for path planning) in a faster way.
Figure~\ref{fig:perroravg} highlights that the accuracy of the estimated map does not improve with the number of AVs: this is caused by the chosen cooperation protocol, where the fusion step favors the latest AV update instead of merging the information from multiple AV updates. This could be overcome by recording the information from multiple updates and estimating the map using a pixel-wise majority voting approach; however, the error probability obtained with a single AV is already sufficiently low (for our path planning purposes), therefore we prefer to save complexity rather than minimizing the error probability.

\subsection{Path Planning Results} 

\begin{figure}[t!]
\centering
\subfloat[]{\includegraphics[width=0.49\columnwidth]{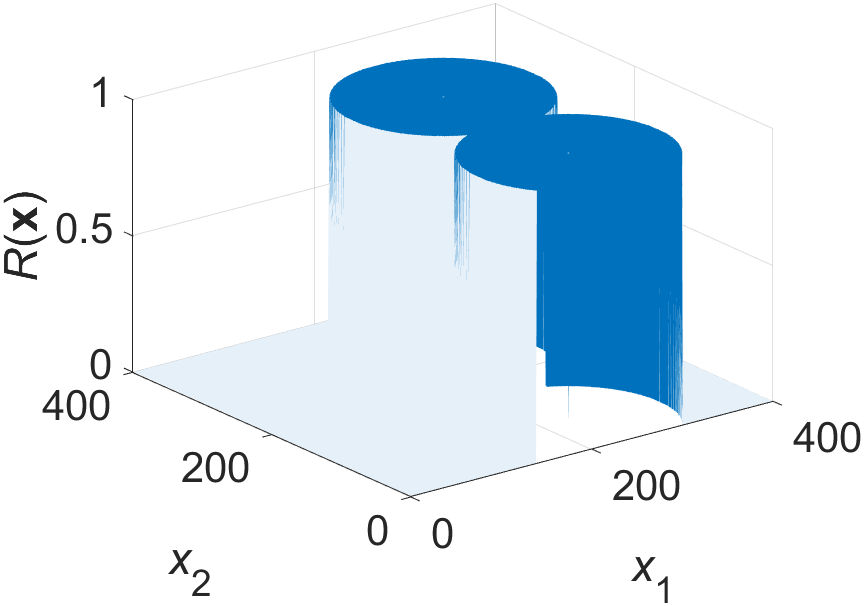}\label{fig:cost1cost}}
\hfill
\subfloat[]{\includegraphics[width=0.49\columnwidth]{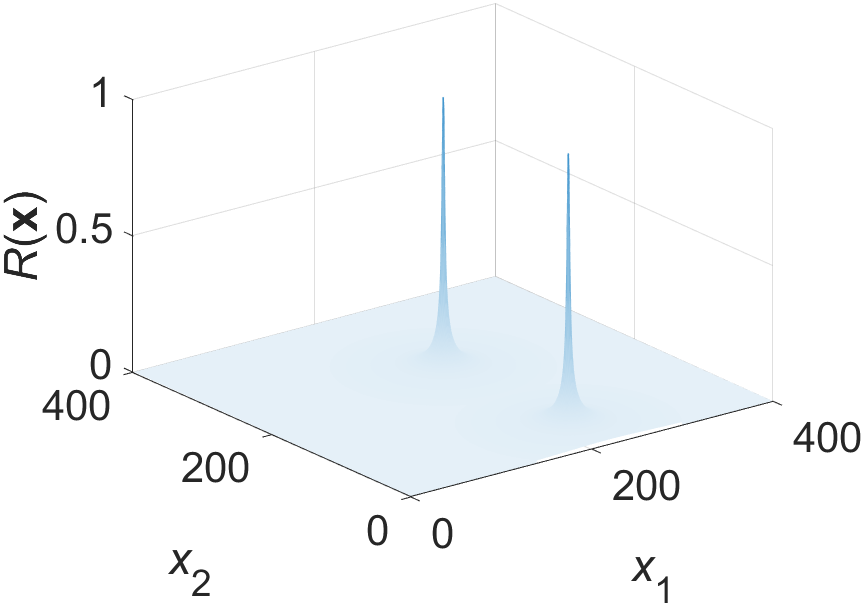}\label{fig:cost2cost}}\\
\subfloat[]{\includegraphics[width=0.49\columnwidth]{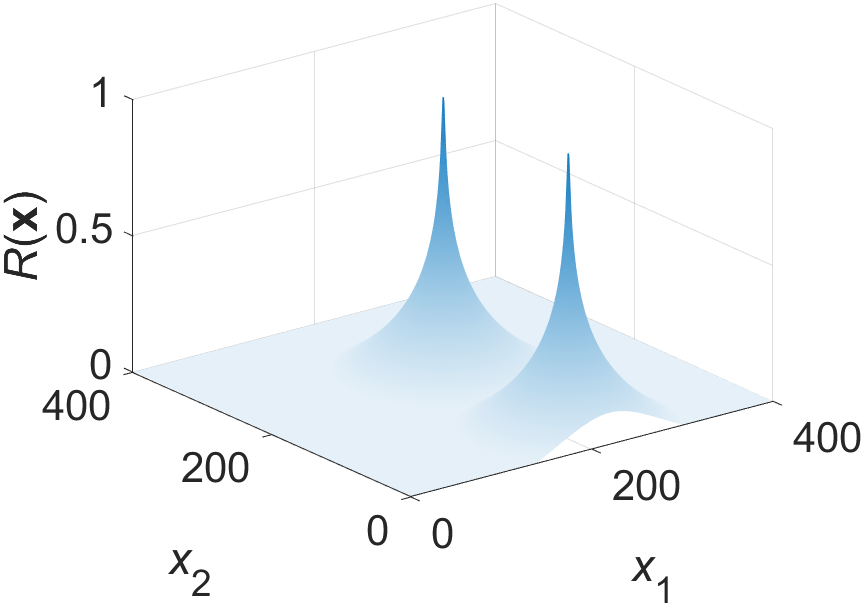}\label{fig:cost3cost}}
\hfill
\subfloat[]{\includegraphics[width=0.49\columnwidth]{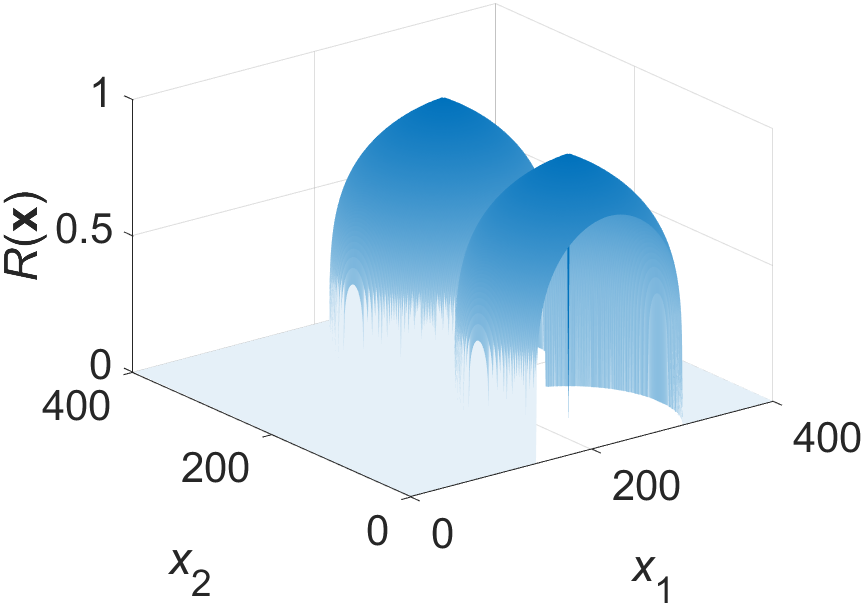}\label{fig:cost4cost}}
\caption{Representation of the on-off (a), amplitude-related (b), capacity-related (c), and tent-shaped (d) weights. \label{fig:costs}}
\end{figure}

To perform path planning, we apply some postprocessing on the estimated obstacle map $\hat{\mathbf{B}}$ obtained during the discovery stage described in Section~\ref{sec:coopmapres}.
Specifically, to obtain the final version $\mathbf{O}$ of the obstacle map, we use a $13\times 13$ Gaussian lowpass filter $\mathbf{h}$ of radius $3$ pixels and an empirically chosen binarization threshold $\tau=0.1$, while the image is kept to the original size of $N\times N = 400\times 400$ pixels ($L=1$). Additionally, for simplicity, only $2$ APs with the same RAT and radius $D^{\text{(MAX)}}=100$ pixels are considered in the radio map. The exponent of the tent-shaped function is chosen as $\beta=0.2$, while the exponent of the amplitude cost is set at $\gamma=1$.
Figure~\ref{fig:costs} shows the shape of the radio map for the considered radio weights: on-off, amplitude-related, capacity-related, and tent-shaped. We highlight that on-off and tent-shaped weights are flatter than amplitude- and capacity-related weights, which give visible differences in the planned paths. Specifically, flatter surfaces route the trajectory close to the coverage disc borders, since deviations from the minimum-distance path add limited radio benefit. Conversely, more peaked surfaces favor AV trajectories pointing to the AP position, as radio gains increase next to the coverage center.

\begin{figure}[t!]
\centering
\subfloat[]{{\includegraphics[width=0.49\columnwidth]{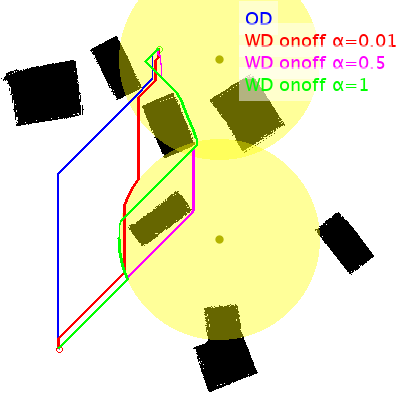}}\label{fig:WD_onoff}}
\hfill
\subfloat[]{{\includegraphics[width=0.49\columnwidth]{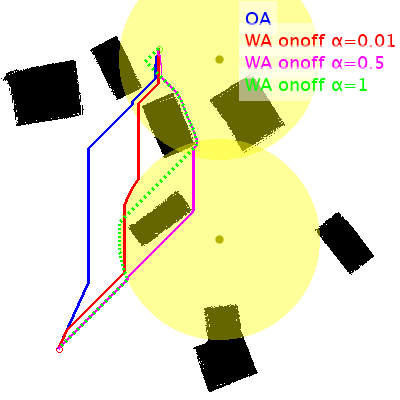}}\label{fig:WA_onoff}}
\caption{Paths with OD, WD (a), WA (b), and on-off weight.
\label{fig:weightedonoff}}
\end{figure}

\begin{figure}[t!]
\centering
\subfloat[]{{\includegraphics[width=0.49\columnwidth]{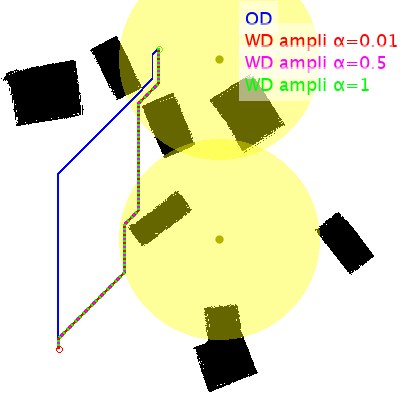}}\label{fig:WD_ampli}}
\hfill
\subfloat[]{{\includegraphics[width=0.49\columnwidth]{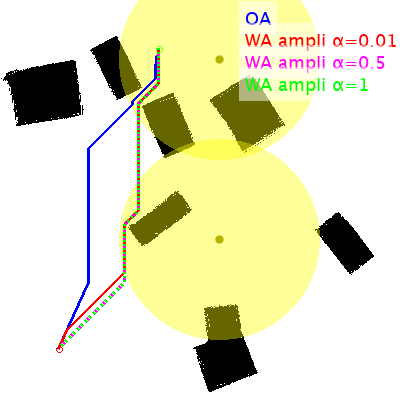}}\label{fig:WA_ampli}}
\caption{Paths with OD, WD (a), WA (b), and amplitude weight.
\label{fig:weightedampli}}
\end{figure}

\begin{figure}[t!]
\centering
\subfloat[]{{\includegraphics[width=0.49\columnwidth]{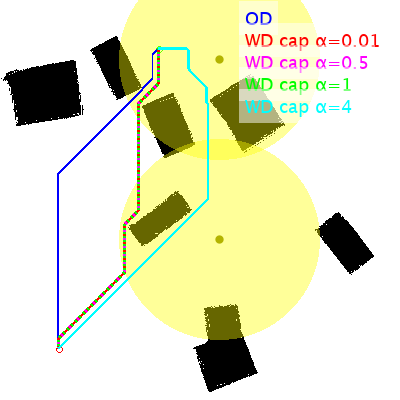}}\label{fig:WD_cap}}
\hfill
\subfloat[]{{\includegraphics[width=0.49\columnwidth]{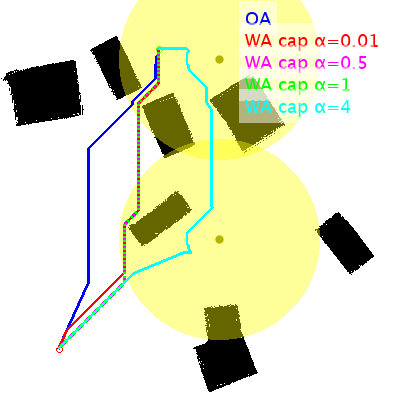}}\label{fig:WA_cap}}
\caption{Paths with OD, WD (a), WA (b), and capacity weight.
\label{fig:weightedcap}}
\end{figure}

\begin{figure}[t!]
\centering
\subfloat[]{{\includegraphics[width=0.49\columnwidth]{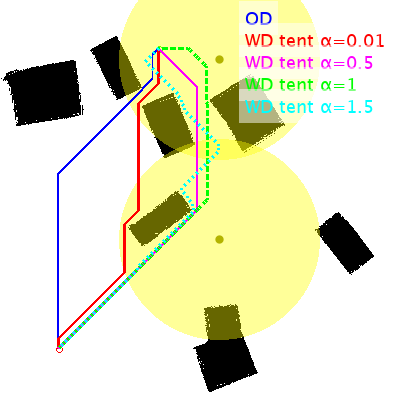}}\label{fig:WD_tent}}
\hfill
\subfloat[]{{\includegraphics[width=0.49\columnwidth]{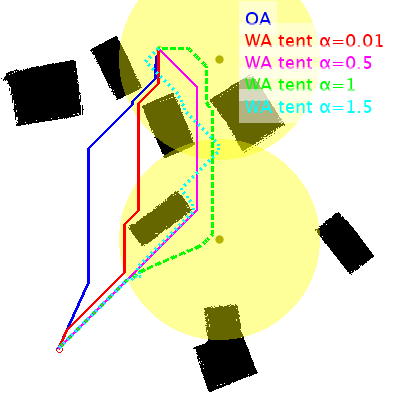}}\label{fig:WA_tent}}
\caption{Paths with OD, WD (a), WA (b), and tent weight.
\label{fig:weightedtent}}
\end{figure}

First, we discuss the results obtained within an invariant scenario, where the starting and ending positions are not changed.
Figures \ref{fig:weightedonoff} through \ref{fig:weightedtent} visually compare the obtained paths adopting the different radio weights, for the proposed WD and WA algorithms, and $\alpha$ varying between $0.01$ and $4$.
When $\alpha$ ranges in $[0,1]$, $\lambda \geq 1/2$ in \eqref{eq:lambdaF1F2}, i.e., the radio experience is not overemphasized with respect to the path distance. However, we also use $\alpha=1.5$ and $\alpha=4$ to test the proposed algorithms on over-the-limit conditions: in such cases, the Dijkstra costs in \eqref{eq:gcumulative} are kept positive or zero by the $\max$ operator.
The position and coverage area of the APs are denoted by the yellow discs, while the small green and red circles indicate the starting and ending position, respectively. We compare the SPP solutions obtained with the existing OD or OA algorithms (both in blue) to the SPP solutions of the proposed weighted algorithms WD or WA (in red, magenta, green or cyan). Note that although OD and OA paths have different shapes, their length $D$ is the same.
We remind that, even when there is line-of-sight between starting and ending points, both OD and OA solutions are not a single straight line, because the AVs can move using specific directions only (such as, multiples of $45$ degrees).

In general, WD and WA produce similar paths for medium-to-low values of $\alpha$, since the WA heuristic is admissible and preserves the optimality of A*. In the simulated cases, the WD/WA paths are longer than their OD/OA counterparts, due to the relevance of the radio term. The differences among the radio surfaces are mainly due to their shapes.
With the on-off weight (Fig.~\ref{fig:weightedonoff}), the planner selects the shortest obstacle-free path that increases the radio-covered portion: the path enters the coverage discs, but obstacles define the final trajectory.
With the amplitude weight (Fig.~\ref{fig:weightedampli}), the steep radio surface renders the path nearly independent of $\alpha$, and WD and WA tend to overlap. 
Instead, the smooth spatial gradient of the capacity (Fig.~\ref{fig:weightedcap}) and tent (Fig.~\ref{fig:weightedtent}) surfaces favors trajectories that penetrate the radio-covered area for large $\alpha$. Visible deviations are produced when $\alpha>1$: WA paths present more direction changes, while WD paths are more regular.

\begin{figure}[t!]
\centering
\subfloat[]{{\includegraphics[width=0.49\columnwidth]{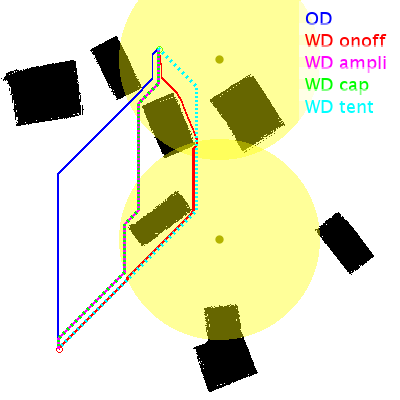}}\label{fig:WD_allweights_050}}
\hfill
\subfloat[]{{\includegraphics[width=0.49\columnwidth]{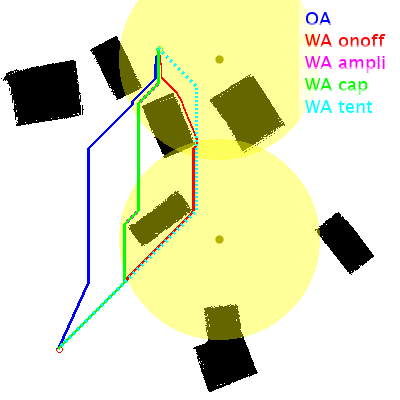}}\label{fig:WA_allweights_050}}
\caption{Paths with OD, WD (a), and WA (b) for different weights and $\alpha=0.5$.
\label{fig:allweights_050}}
\end{figure}

Figure~\ref{fig:allweights_050} compares the paths obtained with $\alpha=0.5$. In this case, WD and WA generate very similar trajectories for all the radio weights, given the optimality preservation of WA. The amplitude and capacity weights result in almost equivalent paths, while the largest differences are obtained by the on-off and tent weights.

\begin{figure}[t!]
\centering
\subfloat[]{{\includegraphics[width=0.49\columnwidth]{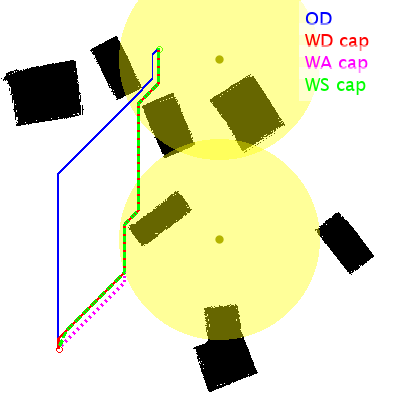}}\label{fig:algs_cap_100_R1}}
\hfill
\subfloat[]{{\includegraphics[width=0.49\columnwidth]{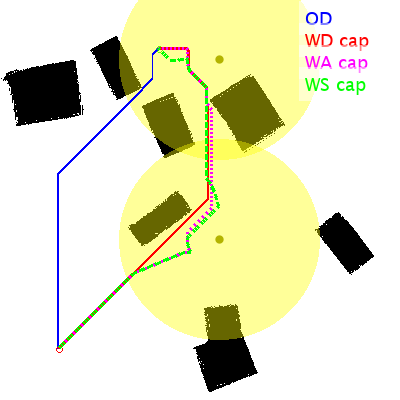}}\label{fig:algs_cap_400_R1}}
\caption{Paths with OD, WD, WA, and WS for $\alpha=1$ (a) and $\alpha=4$ (b).\label{fig:algs_cap_R1}}
\end{figure}

Finally, Figure~\ref{fig:algs_cap_R1} compares WS with OD, WD, and WA for the capacity-weight case. When $\alpha=1$, the weighted algorithms deviate from the OD path and result in substantially coincident trajectories. When $\alpha=4$, the AP attracts the optimal paths, which stay longer inside the coverage regions. In this case, the WS path has visible local deviations with respect to WA and WD, due to its nonclipped heuristic.

Another simulation scenario considers the variation of the starting and ending points. In this case, while the APs and obstacles configurations are kept, the starting and ending positions are randomly changed within the map area, and the resulting metrics are averaged over $500$ runs.

To estimate the actual computational complexity, we consider the running time of the algorithms in MATLAB R2025b on a $3.3$ GHz AMD Ryzen 9 CPU. In our simulations, WD and WA are implemented with array-based data structures and simple frontier management. The asymptotic worst-case complexity of the path planning algorithms is comparable to that of other graph-search methods: for heap-based implementations, it is $\mathcal{O}(N^2\log N)$ for both Dijkstra and A* \cite{elshaer2025,wan2026}. The simulations explore $\alpha \in [0,1]$ and present the results in terms of performance metrics increase with respect to the OA MATLAB implementation.

\begin{figure}
\centering
\includegraphics[width=\columnwidth]{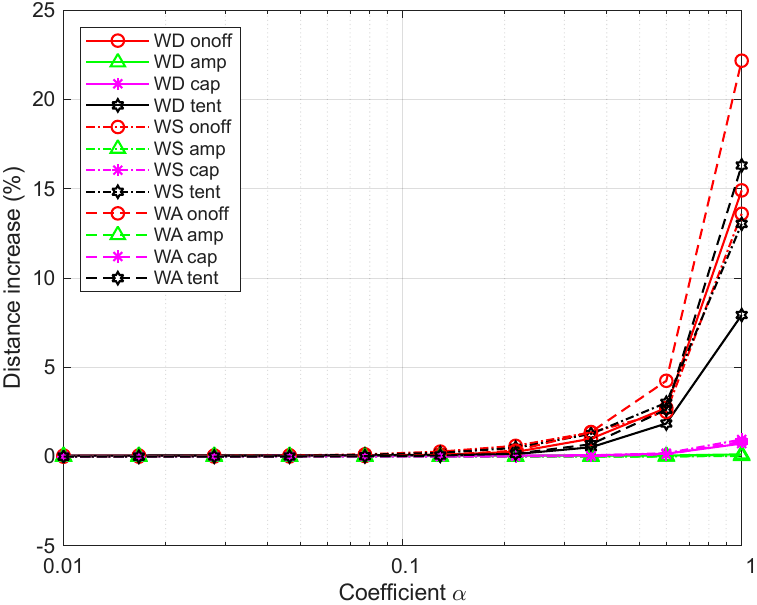}
\caption{Traveled distance increase as function of $\alpha$.\label{fig:distance}}
\end{figure}

Figure~\ref{fig:distance} plots the traveled distance $D$ increase for WD, WA, and WS. When $\alpha \leq 0.2$, all radio weights show no distance increase. For $\alpha>0.2$, the amplitude weight is practically equivalent to the baseline, and the capacity weight increases by less than $1$\% at $\alpha = 1$. On the other hand, the on-off and tent surfaces increase for large $\alpha$: at $\alpha =1$, the on-off case reaches $14$\%--$22$\%, while the tent case ranges from $8$\% to $16$\%. From the simulations, it appears that WA has the greatest distance increase.

\begin{figure}
\centering
\includegraphics[width=\columnwidth]{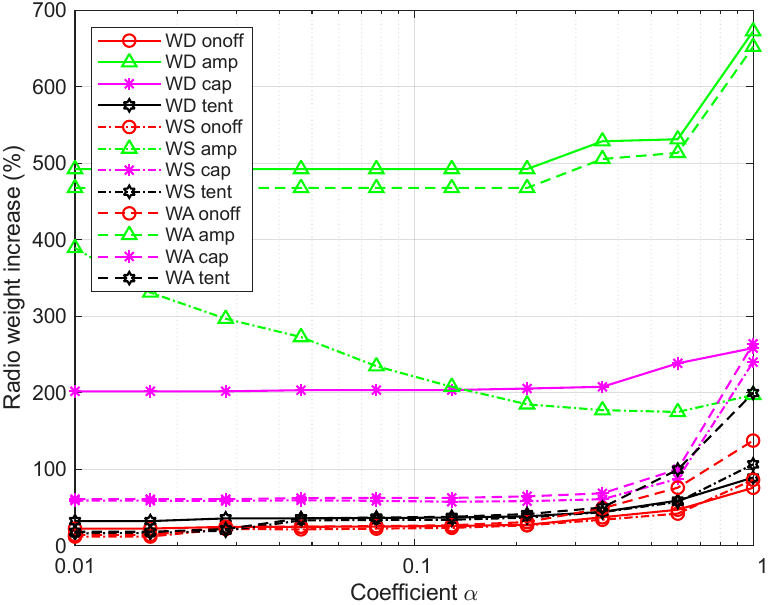}
\caption{Radio weight increase as function of $\alpha$.\label{fig:radiocost}}
\end{figure}

Figure~\ref{fig:radiocost} shows the simulation results for the experienced radio-weight metric $R'$ for WD, WA, and WS. The amplitude weight provides the largest radio increase, from $\approx 200\%$ to $500\%$. The capacity weight, instead, exhibits an algorithm-dependent behavior. For WD, the radio increase stays mostly at $200\%$. By contrast, the WA and WS curves remain around $60\%$--$70\%$ for $\alpha < 0.6$, and increase significantly only when $\alpha \approx 1$, reaching the values of WD.
Also the on-off and tent weights increase noticeably for large $\alpha$, with on-off reaching $80\%$--$140\%$ and tent reaching $100\%$--$200\%$.

\begin{figure}
\centering
\includegraphics[width=\columnwidth]{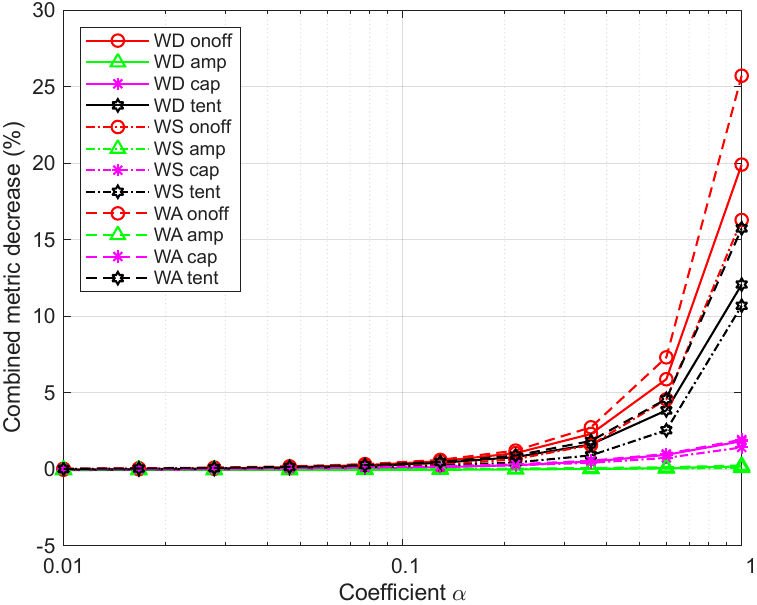}
\caption{Combined metric decrease as function of $\alpha$.\label{fig:combined}}
\end{figure}

Figure~\ref{fig:combined} shows the decrease of the optimization metric $g(\{\mathbf{c}_m\})$ in \eqref{eq:costOD}, which is a weighted combination of the traveled distance and experienced radio weight. The decrease is highest for the on-off and tent weights, but also the capacity weight tends to modify the objective function more significantly. 

\begin{figure}
\centering
\includegraphics[width=\columnwidth]{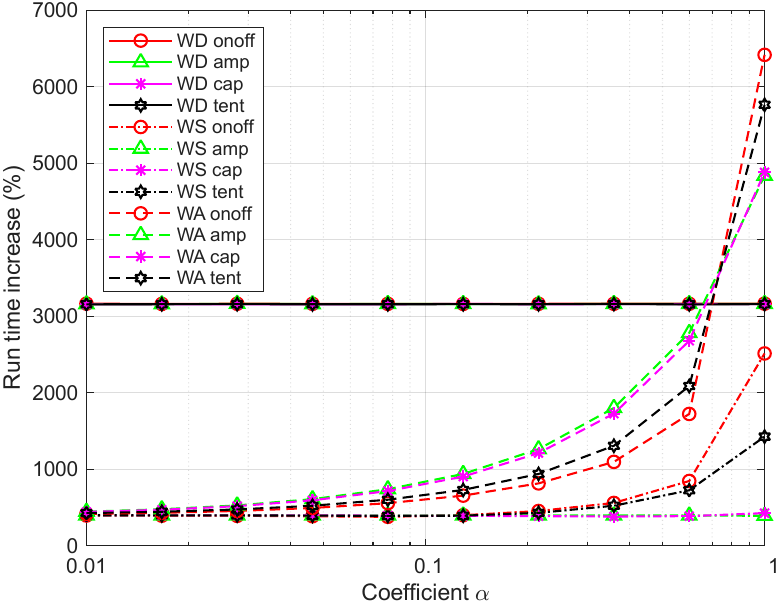}
\caption{Run time increase as function of $\alpha$.\label{fig:complexity}}
\end{figure}

Finally, Figure~\ref{fig:complexity} plots the running time increase with respect to the MATLAB A* implementation, which returns the same optimal path as OD. WD exhibits a constant run-time increase of about $30$ times, independent of $\alpha$ and radio weight. WA and WS are faster for small $\alpha$, but their computational cost depends on the radio surface. In the simulated cases, WA gets slower when $\alpha \approx 1$ and exceeds WD, while WS is still convenient.

Even if other radio models could increase the complexity cost of the radio weight, the proposed planner would not change, keeping the total cost asymptotically similar. A deeper analysis of radio model complexity and computational cost is left to future extensions.

\begin{figure}
\centering
\includegraphics[width=\columnwidth]{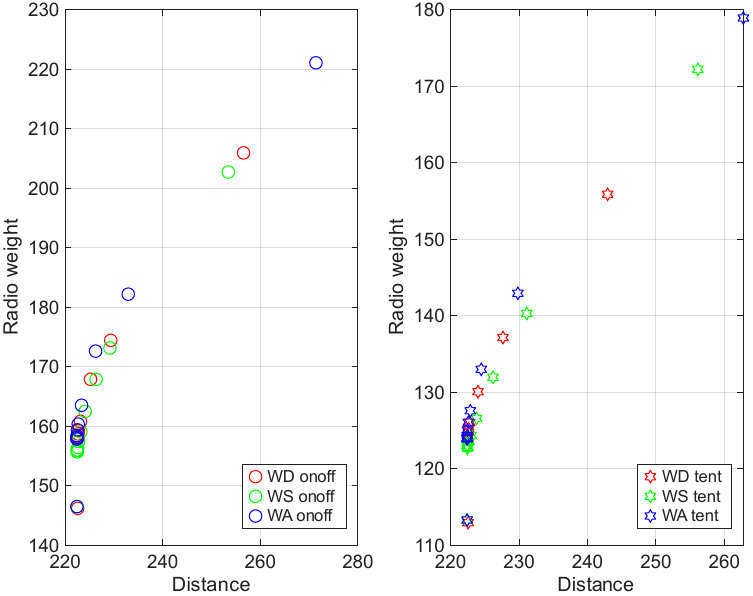}
\caption{Pareto fronts for variable $\alpha$.\label{fig:pareto}}
\end{figure}

Figure~\ref{fig:pareto} shows the Pareto fronts obtained by varying $\alpha$ for the on-off and tent surfaces. The experienced radio coverage is plotted as a function of the path length. The tested values of $\alpha$ are logarithmically spaced in $[0.01,1]$, and also include $\alpha=0$. For both radio surfaces, values of $\alpha < 0.2$ tend to reduce the radio experience more than to increase the path length. For larger $\alpha$, the change of $\alpha$ produces similar variations for both radio metrics.
WD and WA give the best performance, but the WS solutions remain close to the Pareto frontier.
The sensitivity of the proposed algorithms with respect to sensing uncertainty, AP density, coverage radius, and map-processing parameters is outside the scope of the present work.

\begin{figure}
\centering
\includegraphics[width=\columnwidth]{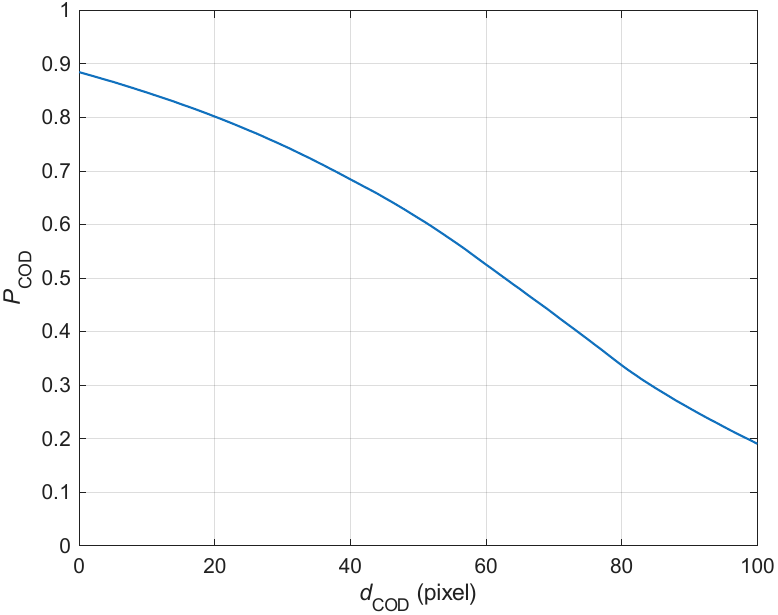}
\caption{Probability of coverage outage for Algorithm 1 of \cite{yang2019}.\label{fig:pcod_vs_dcod_alg1}}
\end{figure}

An alternative strategy for radio coverage aware path planning has been proposed in \cite{yang2019}, which considers a grid-based A* algorithm together with an on-off weight coverage map.
The main difference between our approach and \cite{yang2019} is that our radio weight is inserted in the cost function \eqref{eq:gcumulative}, while the on-off radio weight of \cite{yang2019} appears in the constraints. When the constraints are satisfied, the radio weight in \cite{yang2019} is equivalent to an incremental cost $(1 + \zeta (1 - [\mathbf{R}]_{\mathbf{c}_{m}} )) \Vert \mathbf{c}_m - \mathbf{c}_{m-1} \Vert_2$, which corresponds to our weighted-distance formulation $\max \{ 1 - \alpha [\mathbf{R}]_{\mathbf{c}_{m}}, 0 \}  \Vert \mathbf{c}_m - \mathbf{c}_{m-1} \Vert_2$, up to a positive scaling factor and parameter renormalization. The relation between the two parameters is $\alpha = \zeta / (1+\zeta)$. Since multiplying all single costs by a positive constant does not affect the shortest path, the two formulations are equivalent if the $\max$ operator is not enabled. Our optimization algorithms do not use connectivity constraints and always find a feasible solution, whereas Algorithm 1 of \cite{yang2019} may not find a feasible solution, because of unsatisfied connectivity constraints. 
Specifically, when the on-off weight of Fig.~\ref{fig:weightedonoff} is used, Fig.~\ref{fig:pcod_vs_dcod_alg1} shows the probability $P_{\textrm{COD}}$ that Algorithm 1 of \cite{yang2019} fails to return a solution due to the connectivity outage duration (COD) constraint, as a function of the connection-less distance $d_{\textrm{COD}}$: when $d_{\textrm{COD}} \leq 20$, the probability $P_{\textrm{COD}}$ that Algorithm 1 of \cite{yang2019} is not able to satisfy the COD constraint is quite large, higher than $80$\%. Another advantage of our approach is the support of different radio shapes, while \cite{yang2019} can only use on-off weights. This shows that our formulation generalizes connectivity-aware weighted-distance path planning and does not depend on the specific radio surface.

\section{Conclusion} \label{sec:conclusion}

In this work we have presented novel algorithms to implement radio-aware path planning of an unknown environment by means of cooperative autonomous vehicles employing a simple cooperative mapping strategy. After a cooperative estimation of the obstacle map, the proposed weighted Dijkstra and A* algorithms decide the shortest path between two points not only based on minimum distance, but also with a focus on radio coverage presence and quality, to ensure robust wireless connection towards the data centers. Edge computing can thus take place in the form of task offloading, where the mobile agents execute computationally cumbersome processes on remote edge or cloud servers. Our proposed path planning algorithms are both reasonably accurate and computationally efficient, to be used in all cooperative scenarios where agents can exploit centralized radio access points and computing resources.
We notice that we assumed a constant speed for all the vehicles. In future work, the influence of capacity coverage on data-rate and vehicle speed can be formalized in the problem, and can lead to faster rather than shorter paths.

\end{document}